\documentclass[a4paper,11pt]{article}

\usepackage{jheppub}

\usepackage     [utf8]                  {inputenc}
\usepackage     [T1]                    {fontenc}
\usepackage                             {color}
\usepackage                             {amsmath}
\usepackage                             {amssymb}
\usepackage{bbold}
\usepackage                             {graphicx}
\usepackage     [english]               {babel}
\usepackage                             {hyperref}   
\usepackage                             {float}

\usepackage                             {pdflscape}

\usepackage{array}
\newcolumntype{L}[1]{>{\raggedright\let\newline\\\arraybackslash\hspace{0pt}}m{#1}}
\newcolumntype{C}[1]{>{\centering\let\newline\\\arraybackslash\hspace{0pt}}m{#1}}
\newcolumntype{R}[1]{>{\raggedleft\let\newline\\\arraybackslash\hspace{0pt}}m{#1}}

\usepackage{physics}
\usepackage[makeroom]{cancel}
\usepackage{tabularx}

\definecolor{darkblue}{rgb}{0.0,0.0,0.4}
\definecolor{darkgreen}{rgb}{0.0,0.4,0.0}
\hypersetup{
    colorlinks,
    linkcolor=black,
    citecolor=darkgreen,
    urlcolor=darkblue
}
\usepackage{hyperref}

\definecolor{nicered}{rgb}{0.5,0.,0.}
\definecolor{nicegreen}{rgb}{0.,0.5,0.}
\definecolor{niceblue}{rgb}{0.,0.,0.5}
\hypersetup{
	colorlinks=true,
	linkcolor=niceblue,
	filecolor=nicegreen,      
	urlcolor=niceblue,
	citecolor=nicered,
}

\usepackage{wrapfig}
\usepackage{tikz-feynman}
\allowdisplaybreaks

\newcommand{\0}{$0\nu\beta\beta$}

\newcommand{\Lnuem}{L$\nu$EM}

\title{Positron-Emitting and Electron-Capturing Double-Beta Processes in the Standard Model and Beyond}

\author[a,b]{Luk\'{a}\v{s} Gr\'{a}f}
\emailAdd{lukas.graf@matfyz.cuni.cz}

\author[c,d]{Jenni Kotila}
\emailAdd{jenni.kotila@jyu.fi}

\author[e]{Oliver Scholer}
\emailAdd{scholer@berkeley.edu}

\affiliation[a]{Institute of Particle and Nuclear Physics (IPNP), Faculty of Mathematics and Physics, Charles University
Prague, V Holešovičkách 2, 180 00 Praha 8, Czech Republic}
\affiliation[b]{Institute of Physics, Silesian University in Opava, Bezru{\v{c}}ovo n{\'a}m{\v{e}}st{\'i} 1150/13, 746 01 Opava, Czech Republic}
\affiliation[c]{Finnish Institute for Educational Research, University of Jyv\"{a}skyl\"{a}, P.O. Box 35, 40014, Jyvaskyla, Finland}
\affiliation[d]{International Center for Advanced Training and Research in Physics (CIFRA), 409, Atomistilor Street, Bucharest-Magurele, 077125, Romania}
\affiliation[e]{Department of Physics, University of California Berkeley, CA 94720, USA}

\abstract{We study positron-emitting and electron-capturing double-beta-decay modes as probes complementary to the usual double beta decay. Motivated by the proposed NuDoubt++ experiment, we analyze the candidate isotopes ${}^{78}$Kr, ${}^{106}$Cd, and ${}^{124}$Xe, providing nuclear matrix elements and phase-space factors for both neutrinoful and neutrinoless modes. For the Standard-Model channels, we find that $2\nu\mathrm{ECEC}$ and $2\nu\mathrm{EC}\beta^+$ are the most experimentally accessible, whereas $2\nu\beta^+\beta^+$ remains strongly phase-space suppressed. For the neutrinoless channel, we interpret a projected sensitivity of $T_{1/2}^{0\nu} = 10^{24}$ y in terms of dimension-seven SMEFT operators and find sensitivity to lepton-number-violating new-physics scales of order 1–100 TeV. We further show that measurements in multiple isotopes can help to resolve degeneracies in multi-operator scenarios, making positron-emitting double-beta searches a useful complement to conventional neutrinoless double beta decay experiments.}

\begin{document}

\maketitle

\section{Introduction}
Neutrinoless double beta decay (\0) experiments~\cite{Cirigliano:2022oqy, Agostini:2022zub, GERDA:2020xhi, Augier:2022znx, CUORE:2024ikf, KamLAND-Zen:2024eml, LEGEND:2021bnm, nEXO:2021ujk, CUPID:2022wpt, SNO:2021xpa} are among the most stringent laboratory probes of lepton-number-violating (LNV) physics. A potential observation of this rare process would provide unambiguous proof of beyond-the-Standard-Model (BSM) physics and imply the Majorana nature of neutrinos~\cite{Schechter:1981bd, Takasugi:1984xr}, even though subtleties in the precise interpretation of the experimental signature may involve some caveats~\cite{Duerr:2011zd, Graf:2023dzf}. Indeed, in addition to the widely studied light-neutrino-exchange mechanism (\Lnuem), many BSM models can trigger \0 via distinct mechanisms~\cite{Hirsch:1995ek, Hirsch:1995zi, Hirsch:1996ye, Deppisch:2006hb, Rodejohann:2011mu, Deppisch:2012nb, Graf:2022lhj, Fonseca:2016jbm, Arkani-Hamed:1998wuz, Panella:1997wa, Li:2020flq, Bolton:2021hje, Huang:2013kma}, thus complicating the interpretation of a potential future discovery in terms of a specific particle physics model. Model-independent effective field theory (EFT) approaches to \0~\cite{Pas:1999fc, Pas:2000vn, Deppisch:2012nb, Cirigliano:2017djv, Graf:2018ozy, Cirigliano:2018yza, Deppisch:2020ztt} allow for systematic studies of LNV BSM physics in the context of \0. 

While the double-beta-decay experiments focus on the double-electron-emitting $0\nu\beta^-\beta^-$ decay mode due to its favorable phase space, the recent proposal of the NuDoubt++ collaboration~\cite{NuDoubt:2024jax} to search for double beta decay in positron-emitting/electron-capturing isotopes calls for a closer investigation of these related processes. Such measurements could provide an additional useful input for double-beta nuclear structure calculations as well as potentially help to shed light on the underlying BSM physics that triggers the neutrinoless modes~\cite{Graf:2022lhj}.

In this work, we perform a dedicated EFT analysis of the projected sensitivity of the future NuDoubt++ experiment to both the Standard-Model and BSM modes of these complementary double-beta processes. To this end, we provide a complete set of nuclear matrix elements (NMEs) calculated in the Interacting Boson Model (IBM-2) together with the full set of relevant phase-space factors (PSFs). Specifically, in Section~\ref{sec:2nu} we start by briefly reviewing the two-neutrino-emitting Standard-Model decay modes providing an estimate of the required exposure for an observation of the rarest $2\nu\beta^+\beta^+$ mode. Following this, we turn our focus to the neutrinoless modes. In Section~\ref{sec:eft_formalism} we provide a short introduction to the applied EFT framework. The relevant NMEs and PSFs are presented in Sections~\ref{sec:NMEs} and~\ref{sec:PSFs}, respectively. In Section~\ref{sec:phase_space_observables} we discuss the leptonic phase-space observables in $0\nu\beta^+\beta^+$. Finally, in Section~\ref{sec:BSM_sensitivity} we provide the projected sensitivity to LNV BSM operators in the Standard Model EFT (SMEFT) expected from the upcoming NuDoubt++ experiment.

\section{Neutrinoful Modes}\label{sec:2nu}
\subsection{Half-Life Predictions and Sensitivity}\label{sec:2nusensitivity}
Besides the widely studied two-electron emitting $2\nu\beta^-\beta^-$ decay mode, the Standard Model (SM) allows for three positron-emitting ($\beta^+$) and/or electron-capturing ($\mathrm{EC}$) decay modes related to $2\nu\beta^-\beta^-$ via crossing-symmetry. Equivalently to $2\nu\beta^-\beta^-$, the half-life equation for these modes factorizes into leptonic and nuclear components
\begin{align}
    {T_{1/2}^{2\nu}}^{-1} = g_A^4G_{2\nu}\mathcal{M}_{2\nu},\label{eq:2nuhalflife}
\end{align}
where $G_{2\nu}, \mathcal{M}_{2\nu}$ represent the corresponding isotope and mode-dependent phase-space factor (PSF) and nuclear matrix element (NME), respectively. The relevant PSFs have been calculated e.g.~in Refs.~\cite{Kotila:2013gea, Stoica:2019ajg} by numerically solving the Dirac equation for the exact electron wave-functions in a potential generated by a uniform nuclear charge with electron shell (anti-)screening. In this work, we present a slightly updated formulation in Sec.~\ref{sec:2nuPSFs}. Likewise, the relevant NMEs have been computed in the interacting boson model (IBM-2) with isospin restoration\footnote{Isospin restoration enforces vanishing Fermi NMEs $\mathcal{M}_\mathrm{F}=0$ such that only the Gamow-Teller NME $\mathcal{M}_\mathrm{GT}$ contributes to the total decay amplitude} and under the usual closure approximation (CA) in Ref.~\cite{Barea:2015kwa}. Within these approximations, the full dimensionless NME present in eq.~\eqref{eq:2nuhalflife} can be written as
\begin{align}
    \mathcal{M}_{2\nu} = \frac{m_e}{\Tilde{A}}\mathcal{M}_\mathrm{GT},
\end{align}
where we approximate the closure energy $\Tilde{A}$ as a function of the nuclear mass number $A$ as~\cite{Haxton:1984ggj, Kotila:2013gea}
\begin{align}
    \Tilde{A}\simeq 1.12\,\mathrm{MeV}\times A^{1/2}.
\end{align}
Following Ref.~\cite{Barea:2015kwa}, we apply an effective axial form factor 
\begin{align}
    g_{A,\mathrm{eff}} = 1.269\times A^{-0.18},
\end{align}
to accommodate quenching in $2\nu\beta\beta$ decays. In contrast to Ref.~\cite{Barea:2015kwa}, we keep the axial form factor separated from the NMEs such that rescaling is easily applied by replacing $g_A\rightarrow g_{A,\mathrm{eff}}$ in the half-life equation~\eqref{eq:2nuhalflife}.
In Table~\ref{tab:2nusummary} we summarize the different PSFs calculated using the approach described in Sec.~\ref{sec:2nuPSFs} and NMEs~\cite{Barea:2015kwa} for the neutrino-emitting double beta decay modes in the three NuDoubt++ candidate isotopes \textsuperscript{78}Kr, \textsuperscript{106}Cd, and \textsuperscript{124}Xe as well as the remaining three naturally occurring $2\nu\beta^+\beta^+$ isotopes \textsuperscript{96}Ru, \textsuperscript{130}Ba, and \textsuperscript{136}Ce, together with the corresponding theoretical half-life estimates and, wherever available, the experimentally measured half-life or the lower half-life limit.
\begin{table}[t]
    \centering
    \renewcommand{\arraystretch}{1.25}
    \resizebox{\textwidth}{!}{
    \begin{tabular}{lccccccc}
        \hline\hline
        Isotope
        &
        $\mathcal{M}_{2\nu}$
        &
        $G_{2\nu}^{\beta^+\beta^+}$
        &
        $G_{2\nu}^{\beta^+\mathrm{EC}}$
        &
        $G_{2\nu}^{\mathrm{ECEC}}$
        &
        $T_{1/2}^{2\nu\beta^+\beta^+}$
        &
        $T_{1/2}^{2\nu\beta^+\mathrm{EC}}$
        &
        $T_{1/2}^{2\nu\mathrm{ECEC}}$
        \\
        &
        \cite{Barea:2015kwa}
        &
        $\left[10^{-26}\,\mathrm{yr}^{-1}\right]$
        &
        $\left[10^{-22}\,\mathrm{yr}^{-1}\right]$
        &
        $\left[10^{-21}\,\mathrm{yr}^{-1}\right]$
        &
        $\left[10^{26}\,\mathrm{yr}\right]$
        &
        $\left[10^{22}\,\mathrm{yr}\right]$
        &
        $\left[10^{22}\,\mathrm{yr}\right]$
        \\
        \hline
        $^{78}\mathrm{Kr}$
        &
        $0.190$
        &
        $9.96$
        &
        $3.71$
        &
        $0.61$
        &
        \begin{tabular}{@{}c@{}}
        $1.07$--$24.8$\\
        $(>2.0\times10^{-5})$~\cite{Saenz:1994ty}
        \end{tabular}
        &
        \begin{tabular}{@{}c@{}}
        $2.87$--$66.6$\\
        $(>0.011)$~\cite{Saenz:1994ty}
        \end{tabular}
        &
        \begin{tabular}{@{}c@{}}
        $1.75$--$40.5$\\
        $(1.9)$~\cite{Ratkevich:2017kaz}
        \end{tabular}
        \\
        $^{96}\mathrm{Ru}$
        &
        $0.101$
        &
        $1.07$
        &
        $3.95$
        &
        $2.25$
        &
        \begin{tabular}{@{}c@{}}
        $35.1$--$945$\\
        $(>1.4\times10^{-6})$~\cite{Belli:2013qja}
        \end{tabular}
        &
        \begin{tabular}{@{}c@{}}
        $9.49$--$255$\\
        $(>0.008)$~\cite{Belli:2013qja}
        \end{tabular}
        &
        \begin{tabular}{@{}c@{}}
        $1.67$--$44.9$\\
        (---)
        \end{tabular}
        \\[1.0em]
        $^{106}\mathrm{Cd}$
        &
        $0.114$
        &
        $2.05$
        &
        $6.84$
        &
        $5.11$
        &
        \begin{tabular}{@{}c@{}}
        $14.4$--$416$\\
        $(>1.7\times10^{-4})$~\cite{Belli:2025lsv}
        \end{tabular}
        &
        \begin{tabular}{@{}c@{}}
        $4.32$--$125$\\
        $(>0.077)$~\cite{Belli:2025lsv}
        \end{tabular}
        &
        \begin{tabular}{@{}c@{}}
        $0.578$--$16.7$\\
        $(>0.17)$~\cite{Belli:2025lsv}
        \end{tabular}
        \\[1.0em]
        $^{124}\mathrm{Xe}$
        &
        $0.174$
        &
        $4.92$
        &
        $14.9$
        &
        $16.4$
        &
        \begin{tabular}{@{}c@{}}
        $2.58$--$83.5$\\
        $(>2.0\times10^{-12})$~\cite{Barabash:1989zs}
        \end{tabular}
        &
        \begin{tabular}{@{}c@{}}
        $0.852$--$27.6$\\
        $(>4.8\times10^{-6})$~\cite{Barabash:1989zs}
        \end{tabular}
        &
        \begin{tabular}{@{}c@{}}
        $0.0776$--$2.51$\\
        $(1.1)$~\cite{XENON:2022evz,PandaX-4T:2024fls,LZ:2024wvs}
        \end{tabular}
        \\
        $^{130}\mathrm{Ba}$ 
        &
        0.169
        &
        0.113
        &
        5.73
        &
        14.3
        &
        \begin{tabular}{@{}c@{}}
        $119$--$3993$\\
        (---)
        \end{tabular}
        &
        \begin{tabular}{@{}c@{}}
        $2.35$--$78.6$\\
        (---)
        \end{tabular}
        &
        \begin{tabular}{@{}c@{}}
        $0.094$--$3.15$\\
        $(0.22^*)$~\cite{Meshik:2001ra}
        \end{tabular}
        \\
        $^{136}\mathrm{Ce}$ 
        &
        0.163
        &
        $2.89\times10^{-4}$
        &
        1.86
        &
        11.9
        &
        \begin{tabular}{@{}c@{}}
        $[4.97$--$172]\times10^4$\\
        $(>4.1\times10^{-8})$~\cite{Belli:2017irf}
        \end{tabular}
        &
        \begin{tabular}{@{}c@{}}
        $7.73$--$267$\\
        $(>2.7\times10^{-4})$~\cite{Belli:2017irf}
        \end{tabular}
        &
        \begin{tabular}{@{}c@{}}
        $0.121$--$4.18$\\
        $(>3.2\times10^{-6})$~\cite{Belli:2011zza}
        \end{tabular}
        \\
        \hline\hline
        \end{tabular}
    }
    \caption{
    Nuclear matrix elements~\cite{Barea:2015kwa} computed in the IBM-2 model for the $2\nu\beta^+\beta^+$,
    $2\nu\beta^+\mathrm{EC}$ and $2\nu\mathrm{ECEC}$ modes of the naturally occuring $\beta^+\beta^+$ isotopes $^{78}\mathrm{Kr}$, $^{96}\mathrm{Ru}$, $^{106}\mathrm{Ce}$, $^{124}\mathrm{Xe}$, $^{130}\mathrm{Ba}$, and $^{136}\mathrm{Cd}$, together with the corresponding mode-dependent phase-space factors (see Sec.~\ref{sec:2nuPSFs}).
    The last three columns show the theory-derived half-life ranges for each mode, with the
    experimentally determined half-lives or lower limits given in parentheses. For the theoretical
    half-lives, the lower end of each range corresponds to no quenching, $g_{A,\mathrm{eff}}=g_A$,
    while the upper end corresponds to the quenched value
    $g_{A,\mathrm{eff}}=g_A A^{-0.18}$ estimated in ref.~\cite{Barea:2015kwa}. The quoted half-life for \textsuperscript{130}Ba is marked with an asterisk (*) as an inclusive geochemical derivation in contrast to the other mode-specific laboratory limits.
    }
    \label{tab:2nusummary}
\end{table}

Comparing the half-life estimates provided in Table~\ref{tab:2nusummary} to the expected sensitivity of NuDoubt++~\cite{NuDoubt:2024jax} with a 50\% isotopic \textsuperscript{78}Kr enrichment at $5\,\mathrm{bar}$ overpressure, detection of the positron emitting $2\nu\beta^+\mathrm{EC}$ mode is expected at $7-100\,\mathrm{tonne
\times days}$ exposure. Conversely, detection of the double positron emitting $2\nu\beta^+\beta^+$ mode is outside the projected reach given by the NuDoubt++ collaboration even at $\gtrsim 2000\,\mathrm{tonne\times days}$ exposure~\cite{NuDoubt:2024jax}. 

Assuming the same detection efficiency and background rates apply, detection of the $2\nu\mathrm{ECEC}$ mode requires similar exposure as the $2\nu\mathrm{EC}\beta^+$ mode. However, the physical detection process of these two modes differs largely. While the emitted positron in the $2\nu\mathrm{EC}\beta^+$ mode can deposit $\sim1\,\mathrm{MeV}$ in a detector with a clear $511\,\mathrm{keV}$ annihilation signal, making scintillator setups such as the proposed NuDoubt++ project ideal candidates to probe this mode, the $2\nu\mathrm{ECEC}$ modes primary detection channel is via the de-excitation of the atomic shell by filling the vacant electron holes left after the capture. This de-excitation, typically, releases some $10\,\mathrm{keV}$ of energy via X-rays and Auger electrons~\cite{XENON:2019dti, NuDoubt:2024jax}. This difference in energies and detection channels leads to distinct requirements on the experimental setup and the corresponding energy thresholds. The expected high background rates at these low energies in the first NuDoubt++ iteration complicates a potential detection of the double capture $2\nu\mathrm{ECEC}$ decay.

\begin{figure}[t]
    \centering
    \includegraphics[width=1\linewidth]{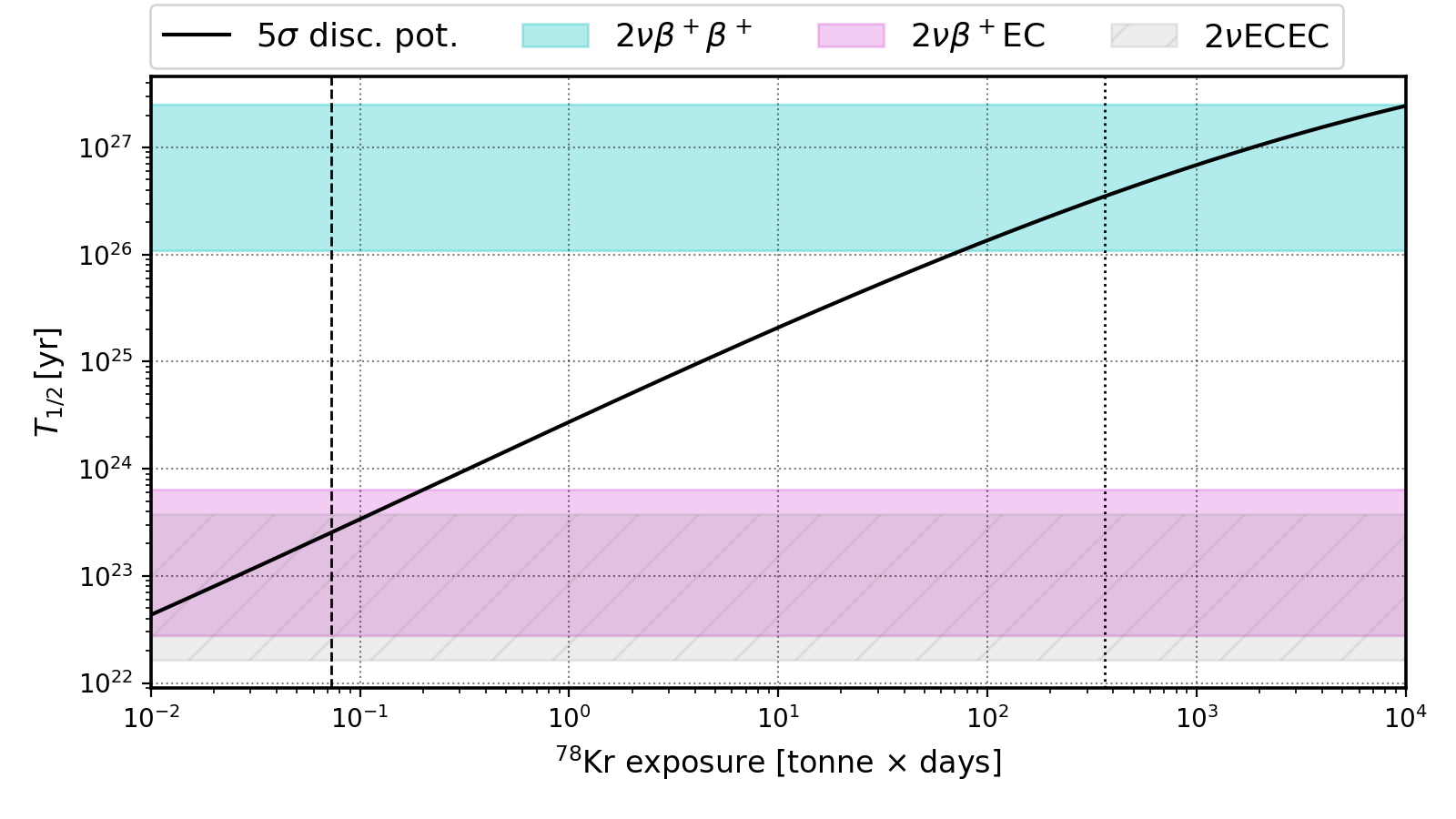}\vspace{-1cm}
    \caption[]{The projected $5\sigma$ discovery potential vs. the theoretical neutrino-emitting double $\beta^+$ and $\mathrm{EC}$ half-lives in a simplified hypothetical low-background NuDoubt++-like \textsuperscript{78}Kr experiment. The horizontal coloured regions showcase the expected theoretical half-life ranges for each decay mode under different $g_A$-quenching assumptions. The two vertical lines represent the first NuDoubt++ \textsuperscript{78}Kr target exposure with a $20\,\mathrm{kg}$ scintillator loaded with $0.2\,\mathrm{kg}$ of \textsuperscript{78}Kr after 2 years of runtime (dashed) and a hypothetical setting with a $10\,\mathrm{ton}$ scintillator loaded with $100\,\mathrm{kg}$ of \textsuperscript{78}Kr and a runtime of 10 years (dotted).
    }
    \label{fig:discovery_potential}
\end{figure}
In order to estimate the required exposure for an observation of the rare double positron emitting $2\nu\beta^+\beta^+$ mode, we perform a naive analysis of the expected discovery potential in a low-background NuDoubt++-like experimental setting. To this end, we assume a simple single-bin counting experiment applying the Feldman-Cousins confidence interval approach~\cite{Feldman:1997qc, Gomez-Cadenas:2010zcc}
with an optimistic background rate of $b\sim1\mathrm{cts}/(\mathrm{yr\times tonne\times ROI})$ with respect to the \textsuperscript{78}Kr isotope mass at a detection efficiency of $50\%$. This background rate roughly translates to the expected background rate in the $0\nu\beta^+\beta^+$ region of interest (ROI) estimated for the NuDoubt++ \textsuperscript{78}Kr setting after applying strict background cuts~\cite{NuDoubt:2024jax}. We define the discovery potential as the required true signal strength (half-life) that has a probability $P_\mathrm{obs}>50\%$ of generating a $5\sigma$ observation. The resulting projected discovery potential vs. \textsuperscript{78}Kr exposure is displayed in Figure~\ref{fig:discovery_potential}. Using this simplified approach, we estimate the required \textsuperscript{78}Kr exposure for a $2\nu\beta^+\beta^+$ detection to be in the range of $10^2-10^4\,\mathrm{tonne \times days}$, depending on the actual $2\nu\beta^+\beta^+$ half-life. The required exposure in terms of scintillator mass can be recovered approximately by multiplying the \textsuperscript{78}Kr exposure by a factor of 100. While this is a largely simplified analysis of the $2\nu\beta^+\beta^+$ discovery potential, we recover the sensitivity profile provided by the NuDoubt++ collaboration in the range of $7-2000\,\mathrm{tonne\times days}$ scintillator exposure within a factor of a few~\cite{NuDoubt:2024jax} and we expect the estimated sensitivity in the higher exposure regimes to be of comparable precision. However, given the optimistic nature of our background estimate which extrapolated the low-background of the $0\nu\beta^+\beta^+$ ROI into the lower energy higher background regime of the full $2\nu\beta^+\beta^+$ spectrum, the true required exposure in a realistic experimental setting may be even higher. On the other hand, a more sophisticated multi-bin 3D analysis may be able to offset a larger background rate in the full $2\nu\beta^+\beta^+$ ROI. Considering the electron capturing $2\nu\beta^+\mathrm{EC}$ and $2\nu\mathrm{ECEC}$ modes, our estimate suggests the discovery potential will cover the full half-life regime at an exposure of $\sim 0.2\,\mathrm{tonne\times days}$ with respect to the \textsuperscript{78}Kr mass supporting the NuDoubt++ projection towards $2\nu\mathrm{EC}\beta^+$ detection.

\subsection{Phase-Space Factors, Energy Spectra and Angular Correlation}\label{sec:2nuPSFs}
The relevant PSFs for the neutrinoful and neutrinoless decay modes are calculated by solving the radial Dirac equation of the electron/positron wavefunctions~\cite{Rose1961RelativisticElectronTheory, Stefanik:2015twa, SALVAT2019165}
\begin{align}
    \Psi(\epsilon,r) = \sum_{\kappa,\mu}\begin{pmatrix}
        g_{\kappa}(\epsilon,r) \chi_\kappa^\mu
        \\
        if_{\kappa}(\epsilon,r) \chi_{-\kappa}^\mu
    \end{pmatrix},
\end{align}
in the Coulomb potential defined by the positively charged nucleus as well as the (anti$-$) screening effects of the surrounding atomic electron cloud. These can be obtained by utilizing the RADIAL algorithm~\cite{SALVAT2019165} which allows us to solve the radial Dirac equations,
\begin{align}
    \frac{\mathrm{d}g_\kappa}{\mathrm{d}r} &= -\frac{\kappa + 1}{r}g_\kappa + (\epsilon + m_e - V(r))f_\kappa,\nonumber
    \\
    \frac{\mathrm{d}f_\kappa}{\mathrm{d}r} &= -(\epsilon-m_e-V(r))g_\kappa + \frac{\kappa-1}{r}f_\kappa,
\end{align}
in the central Coulomb potential $V(r)$ numerically to arbitrary precision. Here, we describe positrons in terms of positive-energy wavefunctions using a sign-flipped potential $V(r)$ which is equivalent to a description in terms of negative-energy wavefunctions without a sign-flip in the potential~\cite{SALVAT2019165}.
Following Refs.~\cite{Kotila:2012zza, Kotila:2013gea, Stefanik:2015twa} we assume a uniform charge distribution inside the nucleus with electron screening parameterized by the appropriate solution to the Thomas-Fermi (TF) function which acts as an effective charge-rescaling $\phi(r)=Z_{\text{eff}}(r)/Z$ screening the visible nuclear charge $Z$ by the negatively charged electron cloud. For a neutral atom, the screening is obtained by solving the TF differential equation
\begin{align}
    \frac{\partial^2\phi}{\partial x^2} = \frac{\phi^{3/2}}{\sqrt{x}},\qquad x=\frac{r}{b},\qquad b=\frac{1}{2}\left(\frac{3\pi}{4}\right)^{2/3}a_0Z^{-1/3},\label{eq:TF}
\end{align}
where $a_0$ is the Bohr radius.
\begin{figure}[t!]
    \centering
    \includegraphics[width=0.9\linewidth]{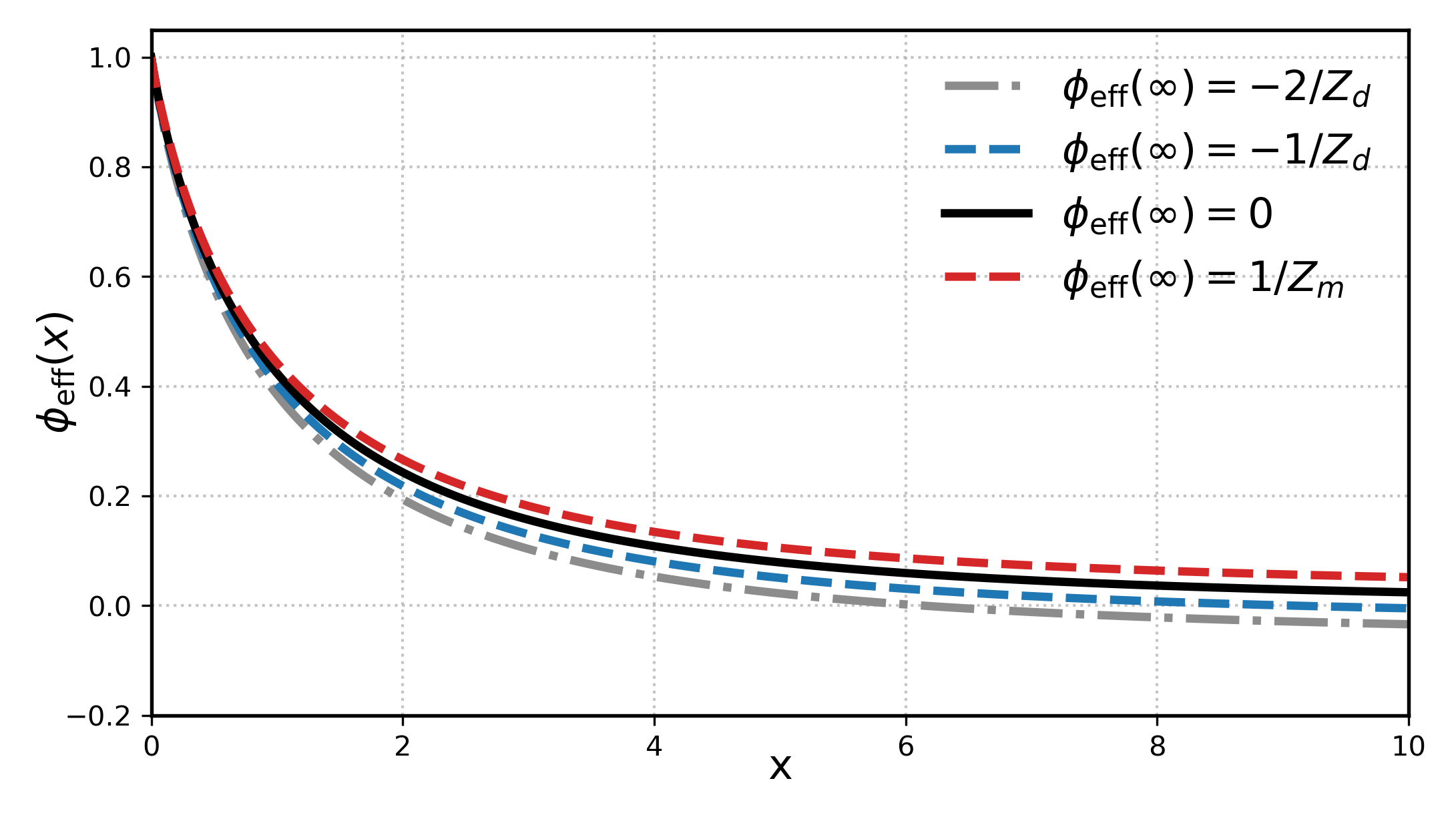}\vspace{-0.5cm}
    \caption{The effective TF screening function obtained using the approximations explained in the main text. The results are displayed for \textsuperscript{78}Kr with $Z_m=36,~Z_d=34$.}
    \label{fig:screening}
\end{figure}
From the interpretation of the Fermi function as an effective charge, we can infer the boundary conditions as being defined by the net charge of the screened nucleus at zero and infinite distance $Z_\infty$
\begin{align}
    \phi(0) = 1,\qquad \phi(\infty) = \frac{Z_\infty}{Z},\qquad Z_\infty=\left\{ \begin{matrix}
        -2, & e^+~\text{in}~\beta^+\beta^+
        \\
        -1 & e^+~\text{in}~\mathrm{EC}\beta^+
        \\
        1 & e^-_b~\text{in}~\mathrm{EC}\beta^+,\mathrm{ECEC}
    \end{matrix} \right.,\qquad Z=\left\{\begin{matrix}
        Z_m, & e^-_b
        \\
        Z_d, & e^+
    \end{matrix}\right.,
\end{align}
where $Z$ is the unscreened charge of the mother nucleus $Z_m$ when considering bound-state electrons $e^-_b$ in the capture modes and the unscreened charge of the daughter nucleus $Z_d$ when considering emitted electrons/positrons $e^\pm$. In contrast to Refs.~\cite{Kotila:2013gea, Stoica:2019ajg}, we take $\phi(\infty)=1$ for all bound state electrons in the initial state atom, following the logic that a single bound state electron $e_b^-$ placed in a neutral atom sees the potential build from the positively charged nucleus with $Z_m$ protons and the remaining $Z_m-1$ electrons in the atomic cloud. In the limit $Z_m\rightarrow1$ this recovers the usual hydrogen atom with a single electron in the cloud that does not feel any screening. The TF equation for the neutral atom with $\phi = Z_\infty/Z = 0$ can be solved via the Majorana method~\cite{Esposito:2001wh}, which we apply here. 
For non-zero boundary conditions $Z_\infty \neq 0$, the simple formulation of the TF presented in eq.~\eqref{eq:TF} does not have a finite real-valued solution ($\phi^{3/2}\notin\mathbb{R}$ for $\phi<0$, $\phi'(x)\rightarrow\infty$ for $\phi(\infty)>0$). Instead, we apply a shifted formulation with the same scaling behavior as the standard TF function
\begin{align}
    \phi_{\mathrm{eff}}(x) = \frac{Z_\infty}{Z} + \frac{Z_e}{Z}\phi_{0}(\Tilde{x}),\qquad \Tilde{x} = x\left(\frac{Z_e}{Z}\right)^{1/3},\qquad Z_e=\left(Z-Z_\infty\right),
\end{align}
which solves the TF-like differential equation,
\begin{align}
    \frac{\mathrm{d}^2\phi_\mathrm{eff}}{\mathrm{d}x^2} = \frac{\left(\phi_\mathrm{eff}-\phi_\mathrm{eff}(\infty)\right)^{3/2}}{\sqrt{x}},\qquad \phi_\mathrm{eff}(0)=1, \qquad \phi_\mathrm{eff}(\infty)=\frac{Z_\infty}{Z} = \frac{Z-Z_e}{Z}.
\end{align}
The rescaled neutral TF function $\phi_0(\Tilde{x})$ corresponds to the screening TF function obtained in a neutral atom with an electron cloud containing $Z_e$ electrons. In Figure~\ref{fig:screening} we display the screening functions $\phi_\mathrm{eff}(x)$ for \textsuperscript{78}Kr.
\begin{figure}[t!]
    \centering
    \includegraphics[width=\linewidth]{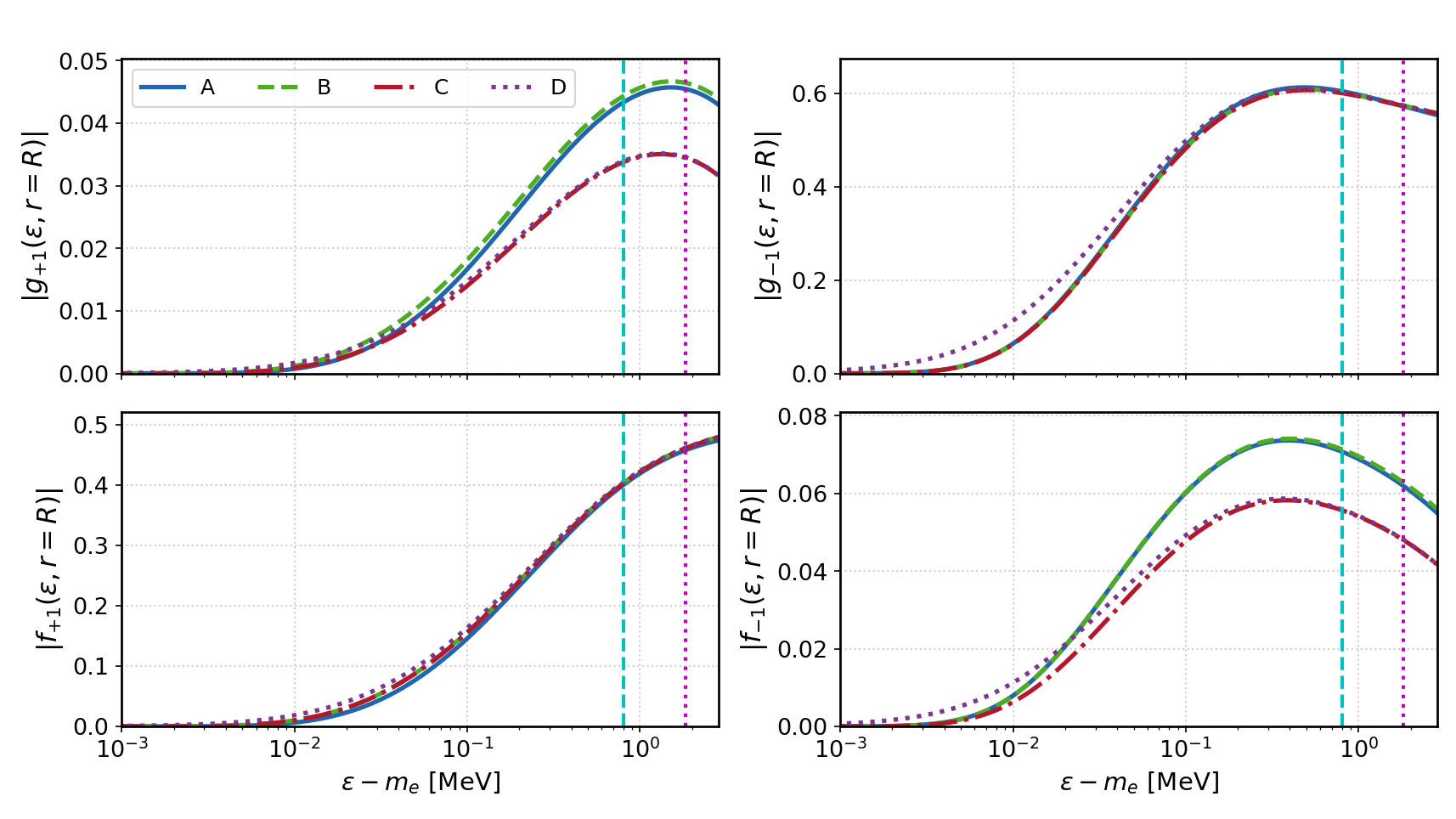}\vspace{-1cm}
    \caption{The energy dependent radial continuum-positron-wavefunctions at the nuclear surface $g_{\kappa}(\epsilon,r=R),~f_{\kappa}(\epsilon,r=R)$ obtained in \textsuperscript{78}Kr with $Z_\infty=-2$ corresponding to the $\beta^+\beta^+$ decay mode. The wavefunctions are dislayed in 4 different approximations~\cite{Stefanik:2015twa}: (A) - leading order approximation to a uniform nuclear charge, (B): exact analytic solution to a point-like nucleus, (C): exact numerical solution to a uniformly charged nucleus, (D): exact numerical solution to a uniformly charged nucleus with electron screening (applied in this work). The $Q-$values of the $\beta^+\beta^+$ and $\mathrm{EC}\beta^+$ decays are displayed as a cyan dashed and magenta dotted vertical line, respectively.}
    \label{fig:wavefunctions}
\end{figure}

The final potential then takes the form
\begin{align}
    V(r) = \pm\phi_\mathrm{eff}(r,Z)\times\left\{\begin{matrix}
        \frac{\alpha Z}{2R}\left(3-\left(\frac{r}{R}\right)^2\right), & r\leq R
        \\
        \frac{\alpha Z}{r}, & r>R
    \end{matrix}\right.,
\end{align}
where the $-$ path describes the attractive potential acting on negatively charged electrons $e^-$ and the $+$ path describes the repulsive potential of the positively charged nucleus acting on the emitted final-state positrons $e^+$. Outside the nuclear radius $r>R$ the screened potential given by this approach can be written as
\begin{align}
    V(r>R) = \pm \alpha\left(\frac{Z}{r} + \frac{Z_e}{r}\left(\phi_0(\Tilde{x})-1\right)\right) = \pm\alpha\left(\frac{Z_\infty}{r} + \frac{Z_e}{r}\phi_0(\Tilde{x})\right),
\end{align}
which simply resembles the potential generated by a nucleus with charge $Z$ screened by an electron cloud with $Z_e$ electrons in a neutral atom. This approach is similar to the screening model adopted by Stoica and Mirea in Ref.~\cite{Stoica:2019ajg}, which takes the form $\phi_\mathrm{eff}^{SM}(r>R) = Z_\infty/Z + Z_e/Z\phi_0(x)$, i.e., without a rescaling of $\phi_0(x)\rightarrow\phi_0(\Tilde{x})$ in the neutral atoms screening function entering $\phi_\mathrm{eff}$, and the resulting screening functions match closely. 

The resulting energy-dependent positron wavefunctions for \textsuperscript{78}Kr evaluated at the nuclear surface $r=R$ are displayed in Figure~\ref{fig:wavefunctions}. We display a comparison of the approximations described above to simpler approaches. The screening effects are most notably in the low-energy regime of $\epsilon-m_e\sim0.001-0.1\,\mathrm{MeV}$, suggesting different levels of relevance for decay modes with continuum single-positron spectra ($2\nu \beta^+\beta^+, 2\nu\mathrm{EC}\beta^+, 0\nu\beta^+\beta^+)$ compared to the fixed-positron-energy decay mode $0\nu\mathrm{EC}\beta^+$ with $\epsilon-m_e\gtrsim1\,\mathrm{MeV}$.

\subsubsection{$2\nu\beta^+\beta^+$}
The phase-space factor of the double positron emitting $2\nu\beta^+\beta^+$ decay follows the same structure as the usual $2\nu\beta^-\beta^-$ mode. It can be written as~\cite{Kotila:2012zza}
\begin{align}
    G_{2\nu}^{\beta\beta} = \frac{G_F^4 V_{ud}^4}{64\pi^7 m_e^2}\int\big(g_{2\nu}(\epsilon_1,\epsilon_2, R) + h_{2\nu}(\epsilon_1,\epsilon_2, R)\cos\theta\big)\omega_1^2\omega_2^2k_1k_2\epsilon_1\epsilon_2\mathrm{d}\omega_1\,\mathrm{d}\epsilon_1\,\mathrm{d}\epsilon_2\,\mathrm{d}\cos\theta,
\end{align}
where $\epsilon_i, k_i$ are the energy and momentum of the $i$-th emitted positron, $\omega_i$ represents the energy of the $i$-th neutrino, and $\theta$ is the opening angle between the two emitted positrons. The integral is taken over the full kinematically allowed phase-space,
\begin{align}
    \int \mathrm{d}\omega_1\,\mathrm{d}\epsilon_1\,\mathrm{d}\epsilon_2\,\mathrm{d}\cos\theta = \int_{m_e}^{Q_{\beta\beta}+m_e}\mathrm{d}\epsilon_1 \int_{m_e}^{Q_{\beta\beta}+2m_e - \epsilon_1}\mathrm{d}\epsilon_2 \int_{0}^{Q_{\beta\beta}+ 2m_e - \epsilon_1-\epsilon_2}\mathrm{d}\omega_1\int_{-1}^1\mathrm{d}\cos\theta,
\end{align}
and the functions $g_{2\nu}$ and $h_{2\nu}$ parameterize the absolute magnitude and the angular dependency of the phase-space, respectively. They depend on the radial wavefunctions $g_\kappa,f_\kappa$ as~\cite{Kotila:2012zza}
\begin{align}
    g_{2\nu} &= \frac{1\Tilde{A}^2}{3\log2}\Big[\big|f^{-1-1}\big|^2 + \big|f_{11}\big|^2 + \big|{f^{-1}}_{1}\big|^2 + \big|{f_{1}}^{-1}\big|^2\Big]\left[\left<K_N\right>^2 + \left<L_N\right>^2 + \left<K_N\right>\left<L_N\right>\right],\nonumber
    \\
    h_{2\nu} &= -\frac{4\Tilde{A}^2}{9\log2}\mathrm{Re}\Big[f^{-1-1}f_{11}^* + {f^{-1}}_{1}{f_1}^{-1*}\Big]\left[2\left<K_N\right>^2 + 2\left<L_N\right>^2 + 5\left<K_N\right>\left<L_N\right>\right],
\end{align}
with
\begin{align}
    \left<K_N\right> &= \frac{1}{\epsilon_1+\omega_1+\left<E_N\right>-E_I} + \frac{1}{\epsilon_2+\omega_2+\left<E_N\right>-E_I},\nonumber
    \\
    \left<L_N\right> &= \frac{1}{\epsilon_1+\omega_2+\left<E_N\right>-E_I} + \frac{1}{\epsilon_2+\omega_1+\left<E_N\right>-E_I},
\end{align}
and the radial wavefunctions $g_\kappa,f_\kappa$ evaluated at the nuclear radius $R$ entering via
\begin{align}
    &f^{-1-1} = g_{-1}(\epsilon_1,R)g_{-1}(\epsilon_2,R), & &f_{11}=f_1(\epsilon_1,R)f_1(\epsilon_2,R),\nonumber
    \\
    &{f^{-1}}_1 = g_{-1}(\epsilon_1,R)f_1(\epsilon_2,R), & &{f_1}^{-1}=f_1(\epsilon_1,R)g_{-1}(\epsilon_2,R).
\end{align}
The energy difference between the intermediate and initial state $E_N-E_I$ is averaged via
\begin{align}
    \Tilde{A} = \frac{1}{2}W_0 + \left<E_N\right>-E_I \simeq 1.12\times A^{1/2}\,\mathrm{MeV},\qquad W_0 = Q_{\beta\beta}+2m_e.
\end{align}
The $Q-$value of the $\beta^+\beta^+$ mode is given in terms of the mass difference of the initial and final state neutral atoms $\Delta M$ as
\begin{align}
    Q_{\beta\beta} = \Delta M - 4m_e.
\end{align}

\subsubsection{$2\nu\mathrm{EC}\beta^+$}
The PSF of the mixed capture and emission mode $2\nu\mathrm{EC}\beta^+$ is given by~\cite{Kotila:2013gea}
\begin{align}
    G_{2\nu}^{\mathrm{EC}\beta^+} = \frac{G_F^4V_{ud}^4 \Tilde{A}^2 m_e}{24\log(2)\pi^5}\sum_n \mathcal{B}_{n,-1}^2 \int_{m_e}^{Q_{\mathrm{EC}\beta}-\epsilon_{b,n} + m_e}\mathrm{d}\epsilon_{p}\int_0^{Q_{\mathrm{EC}\beta}-\epsilon_{b,n} - \epsilon_{p} + m_e} \mathrm{d}\omega_1 \nonumber
    \\
    \times \big[(g_{-1}^2(\epsilon_{p}, R) + f_1^2(\epsilon_p, R)\big]\Big[\left<K_N\right>^2 + \left<L_N\right>^2 + \left<K_N\right>\left<L_N\right>\Big]\omega_1\omega_2k_p\epsilon_p,
\end{align}
where $\epsilon_{b,n}$ denotes the (positive) binding energy of the $S1/2$ state in the $n$-th shell which we extract numerically using the RADIAL algorithm~\cite{SALVAT2019165}. 

The probability of a bound-state electron in the $n-$th shell being found at the nuclear radius $R$ is given by~\cite{Kotila:2013gea, Doi:1991xf}
\begin{align}
    \mathcal{B}^2_{n,\kappa} = \frac{1}{4\pi m_e^3 R^2 a_0}\left[(g_{n,\kappa}^{b}(R)^2 + f_{n,\kappa}^{b}(R)^2\right].
\end{align}
Here, we restrict our analysis to the leading order contributions from the $K,n=0$ and $L_I,n=1$ shells with $\kappa=-1$.

\subsubsection{$2\nu\mathrm{ECEC}$}
In the double electron capture mode, two bound state electrons of the initial-state atom are captured by the nucleus emitting two neutrinos in the process. As discussed in Sec.~\ref{sec:2nusensitivity}, experimental detection of this decay mode differs significantly from the positron emitting modes. The PSFs are calculated in analogy to the $2\nu\mathrm{EC}\beta^+$ scenario by performing the kinematic integrals over the neutrino phase-space weighted by the electron capture probabilities $\mathcal{B}^2_{n,\kappa}$ as~\cite{Doi:1991xf, Kotila:2013gea}
\begin{align}
    G_{2\nu}^\mathrm{ECEC} = 
    \frac{\Tilde{A}^2 G_F^4V_{ud}^4m_e^4}{24\log(2)\pi^3}\sum_{n,m}\mathcal{B}^2_{n,-1}\mathcal{B}^2_{m,-1}
    \int_0^{Q_\mathrm{ECEC}-\epsilon_{b_n} - \epsilon_{b_m}} \nonumber
    \\
    \times\Big[\left<K_N\right>^2 + \left<L_N\right>^2 + \left<K_N\right>\left<L_N\right>\Big]\omega_1\omega_2 \mathrm{d}\omega_1.
\end{align}

\subsubsection{Spectra and Angular Correlation}

\begin{figure}[t!]
    \centering
    \includegraphics[width=1\linewidth]{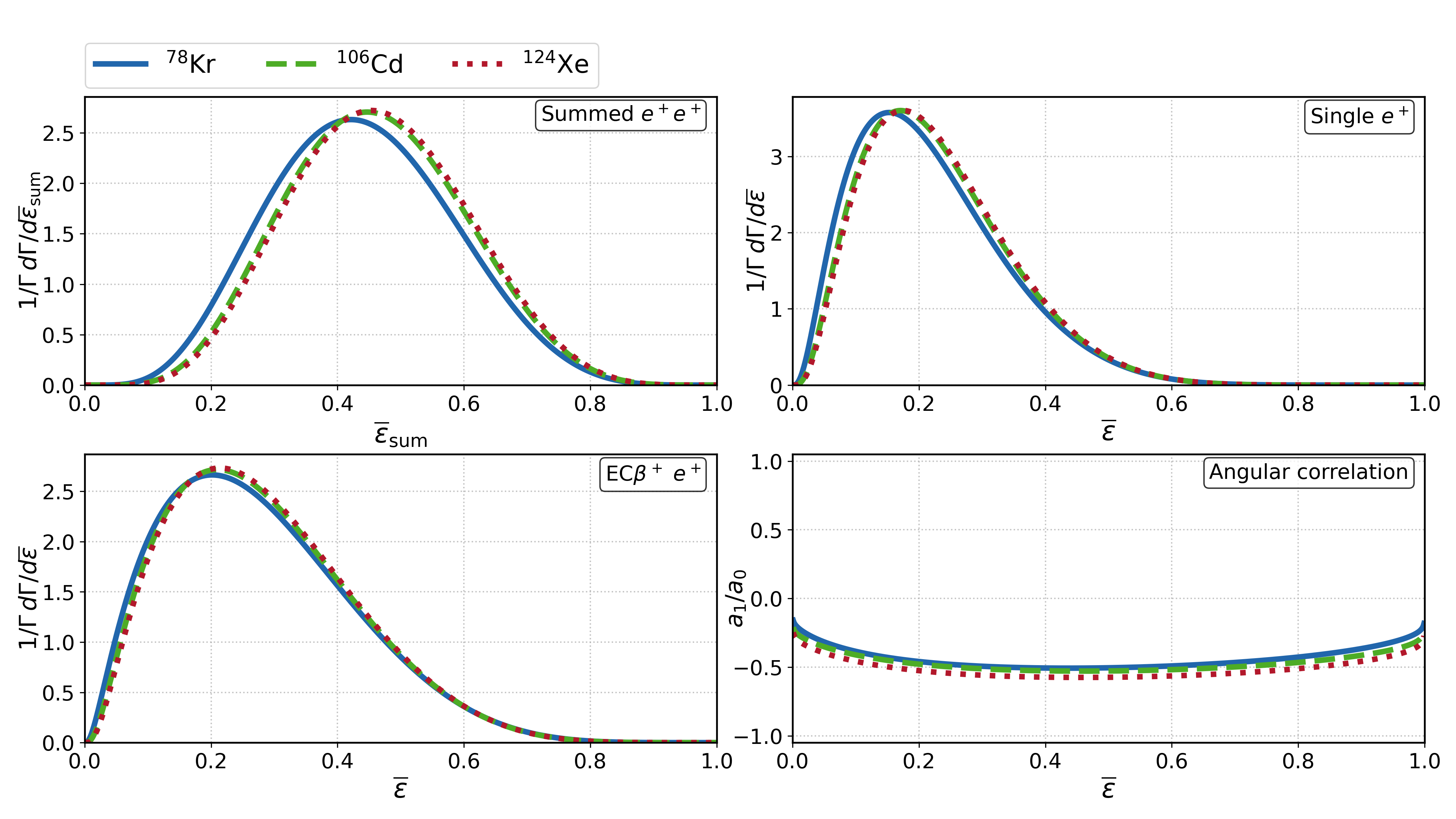}\vspace{-1cm}
    \caption{Normalized summed positron spectra (upper left), single positron spectra (upper right), and angular correlations (lower right) for $2\nu\beta^+\beta^+$ in the three NuDoubt++ candidate isotopes \textsuperscript{78}Kr, \textsuperscript{106}Cd, and \textsuperscript{124}Xe. Lower left panel depicts the positron energy spectrum for $2\nu EC \beta^+$ in the same three isotopes.}
    \label{fig:spectra_2nu}
\end{figure}

The normalized single and summed positron spectra for the SM two-neutrino modes are easily obtained by taking the appropriate derivative of the PSFs,
\begin{align}
    \frac{1}{\Gamma}\frac{\mathrm{d}\Gamma}{\mathrm{d}\epsilon_1}
    = \frac{1}{G_{2\nu}}\frac{\mathrm{d}G_{2\nu}}{\mathrm{d}\epsilon_1},
    \qquad
    \frac{1}{\Gamma}\frac{\mathrm{d}\Gamma}{\mathrm{d}\epsilon_\mathrm{sum}}
    = \frac{1}{G_{2\nu}}\frac{\mathrm{d}G_{2\nu}}{\mathrm{d}\epsilon_\mathrm{sum}},
    \qquad
    \epsilon_\mathrm{sum} = \epsilon_1 + \epsilon_2.
\end{align}
Obviously, for the $2\nu\mathrm{EC}\beta^+$ mode with a single emitted positron, the summed and single positron spectra are identical with $\epsilon_1=\epsilon_p$ and $\epsilon_2=0$. Similarly, for the double positron emitting decay we can define the angular correlation as
\begin{align}
    \frac{a_1}{a_0} = \frac{\mathrm{d}H^{\beta\beta}_{2\nu}/\mathrm{d}\epsilon_1}{\mathrm{d}G^{\beta\beta}_{2\nu}/\mathrm{d}\epsilon_1},\qquad \frac{\mathrm{d}\Gamma}{\mathrm{d}\cos\theta~\mathrm{d\epsilon_1}} = a_0 + a_1\cos\theta,
\end{align}
where we used the integrated angular PSF,
\begin{align}
    H_{2\nu}^{\beta\beta} = \frac{G_F^4 V_{ud}^4}{64\pi^7 m_e^2}\int\big(h_{2\nu}(\epsilon_1,\epsilon_2, R)\cos\theta\big)\omega_1^2\omega_2^2k_1k_2\epsilon_1\epsilon_2\mathrm{d}\omega_1\,\mathrm{d}\epsilon_1\,\mathrm{d}\epsilon_2\,.
\end{align}
In Figure~\ref{fig:spectra_2nu} we present the normalized single and summed positron spectra as well as the angular correlations of the $2\nu\beta^+\beta^+$ decay mode. The plots are displayed in dependence on the normalized kinetic positron energies,
\begin{align}
    \overline{\epsilon} = \frac{\epsilon_{1}-m_e}{Q},\qquad \overline{\epsilon}_\mathrm{sum} = \frac{\epsilon_1+\epsilon_2-2m_e}{Q},
\end{align}
where $\epsilon_i$ is the energy of the $i$-th positron. All three leptonic phase-space observables display only minor to nearly vanishing isotope dependence due to the SM supplied helicity constraints and similar Q-values.

\section{Neutrinoless Modes}
\subsection{Effective Lepton-Number-Violating Interactions}\label{sec:eft_formalism}
\begin{table}[t]
\centering
\renewcommand{\arraystretch}{1.25}
\begin{tabular}{llll}
\hline\hline
Class & Structure & Operator & Definition \\
\hline
$1$ & $\psi^2 H^4$
&
$\mathcal{O}^{(7)}_{LH}$
&
$\epsilon_{ij}\epsilon_{mn}
\left(L_i^T C L_m\right) H_j H_n \left(H^\dagger H\right)$
\\
\hline
$2$ & $\psi^2 H^2 D^2$
&
$\mathcal{O}^{(7)}_{LHD1}$
&
$\epsilon_{ij}\epsilon_{mn}
\left(L_i^T C (D_\mu L)_j\right) H_m (D^\mu H)_n$
\\
&
&
$\mathcal{O}^{(7)}_{LHD2}$
&
$\epsilon_{im}\epsilon_{jn}
\left(L_i^T C (D_\mu L)_j\right) H_m (D^\mu H)_n$
\\
\hline
$3$ & $\psi^2 H^3 D$
&
$\mathcal{O}^{(7)}_{LHDe}$
&
$\epsilon_{ij}\epsilon_{mn}
\left(L_i^T C\gamma_\mu e\right) H_j H_m (D^\mu H)_n$
\\
\hline
$4$ & $\psi^2 H^2 X$
&
$\mathcal{O}^{(7)}_{LHB}$
&
$\epsilon_{ij}\epsilon_{mn} g'
\left(L_i^T C\sigma^{\mu\nu} L_m\right) H_j H_n B_{\mu\nu}$
\\
&
&
$\mathcal{O}^{(7)}_{LHW}$
&
$\epsilon_{ij}(\epsilon\tau^I)_{mn} g_2
\left(L_i^T C\sigma^{\mu\nu} L_m\right) H_j H_n W^I_{\mu\nu}$
\\
\hline
$5$ & $\psi^4 D$
&
$\mathcal{O}^{(7)}_{LLduD1}$
&
$\epsilon_{ij}
\left(d\gamma_\mu u\right)
\left(L_i^T C(D^\mu L)_j\right)$
\\
\hline
$6$ & $\psi^4 H$
&
$\mathcal{O}^{(7)}_{LLeH}$
&
$\epsilon_{ij}\epsilon_{mn}
\left(e L_i\right)\left(L_j^T C L_m\right)H_n$
\\
&
&
$\mathcal{O}^{(7)}_{LLQdH1}$
&
$\epsilon_{ij}\epsilon_{mn}
\left(d L_i\right)\left(Q_j^T C L_m\right)H_n$
\\
&
&
$\mathcal{O}^{(7)}_{LLQdH2}$
&
$\epsilon_{im}\epsilon_{jn}
\left(d L_i\right)\left(Q_j^T C L_m\right)H_n$
\\
&
&
$\mathcal{O}^{(7)}_{LLQuH}$
&
$\epsilon_{ij}
\left(Q_m u\right)\left(L_m^T C L_i\right)H_j$
\\
&
&
$\mathcal{O}^{(7)}_{LeudH}$
&
$\epsilon_{ij}
\left(L_i^T C\gamma_\mu e\right)
\left(d\gamma^\mu u\right)H_j$
\\
\hline\hline
\end{tabular}
\caption[Lepton-number-violating $\Delta L=2$ operators at SMEFT dimension 7.]{
Lepton-number-violating $\Delta L=2$ operators at SMEFT dimension 7.
Table adapted from Ref.~\cite{Cirigliano:2017djv}.
}
\label{tab:SMEFT_Dim_7}
\end{table}
Turning to the neutrinoless decay modes, a general EFT framework for the calculation of \0 decay rates and phase-space observables has been derived in Refs.~\cite{Cirigliano:2017djv, Cirigliano:2018yza}. While originally developed to describe the $0\nu\beta^-\beta^-$ decay, this EFT framework can be readily adapted towards other double beta processes related via crossing-symmetry. The calculation is performed by following a chain of EFTs starting with LNV operators in the $SU(3)_C\times SU(2)_L\times U(1)_Y$-invariant SMEFT~\cite{Grzadkowski:2010es, Lehman:2014jma, Henning:2015alf, Liao:2020jmn, Li:2020xlh},
\begin{align}
    \mathcal{L}_\mathrm{SMEFT} = \mathcal{L}_\mathrm{SM} + \sum_{d\geq5}\sum_{i}C_i^{(d)}\mathcal{O}_i^{(d)}.
\end{align}
At SMEFT level, LNV appears only at odd operator dimensions~\cite{Kobach:2016ami} with the neutrino mass generating Weinberg operator at dimension 5~\cite{Weinberg:1979sa} and 12 different $\Delta L=2$ operators at dimension-7 SMEFT. In what follows we will primarily focus on more exotic \0 mechanisms induced via dimension-7 SMEFT operators and their imprints in positron emitting $0\nu\beta^+\beta^+$ and $0\nu\mathrm{EC}\beta^+$ decays that could be potentially used for pinpointing some of these non-standard LNV sources. At the same time, while EFT power counting would suggest that lower-dimensional operators should give the leading contributions, this ordering does not need to hold in realistic UV-completions, where lower-dimensional operators can be suppressed by loop factors, flavour structure, or chirality constraints~\cite{Fridell:2024pmw, Esser:2026ehc}. A comprehensive study of complementary probes of LNV at dimension-7 in SMEFT has been performed in Ref.~\cite{Fridell:2023rtr}. In Table~\ref{tab:SMEFT_Dim_7} we present the 12 LNV $\Delta L=2$ SMEFT dimension-7 operators that contribute to \0 decay at tree or 1-loop level.

Following along the EFT chain formalism, the SMEFT operators are subsequently matched onto an $SU(3)_C\times U(1)_\mathrm{EM}$ invariant low-energy EFT (LEFT) \cite{Jenkins:2017jig, Liao:2020zyx} valid below the scales of electroweak symmetry breaking and, finally, to chiral perturbation theory~\cite{Gasser:1983yg} ($\chi$PT) and chiral EFT~\cite{Weinberg:1991um} to derive the relevant nuclear operators in the non-perturbative regime of QCD. The detailed calculations including the matching procedures and RGE evolution~\cite{Antusch:2001ck, Jenkins:2013zja, Jenkins:2013wua, Alonso:2013hga, Zhang:2023kvw, Zhang:2023ndw, Cirigliano:2018yza, Jenkins:2017dyc} between the relevant energy scales have been automated in the Python tool $\nu$DoBe~\cite{Scholer:2023bnn}, an upgraded, in-development version of which we utilize for this work.

Similar to the $2\nu\beta\beta$ half-life described in eq.~\eqref{eq:2nuhalflife}, the general $0\nu\beta\beta$ half-life formula can be written in terms of leptonic PSFs $G_{0k}$ and nuclear sub-amplitudes $\mathcal{A}_{k}(C_i)$ that depend on the precise Wilson coefficients $C_i$ of the LNV model studied as well as the relevant NMEs and low-energy constants (LECs),
\begin{align}
    T^{-1}_{1/2} = g_A^4 \sum_k G_{0k} \left| \mathcal{A}_k \left( C_i \right)\right|^2.\label{eq:0nuhalflife}
\end{align}
Unfortunately, the numerical values of some relevant LECs are currently unknown. Although their magnitude can be estimated from a naive dimensional analysis~\cite{Manohar:1983md, Jenkins:2013sda}, we simply set it to zero in our current work to provide a more robust and conservative analysis.
The complete framework and explicit formulation are described in Refs.~\cite{Cirigliano:2017djv, Cirigliano:2018yza}.

\subsection{Nuclear Matrix Elements}\label{sec:NMEs}
Crucially, with the exception of $M_T^{AA}$, the general \0 half-life equation derived in Refs.~\cite{Cirigliano:2017djv, Cirigliano:2018yza} depends on the same NMEs as the widely studied light- and heavy-neutrino exchange mechanisms. Therefore, for the standard $0\nu\beta^-\beta^-$ modes, the relevant NMEs are readily available in existing literature within multiple nuclear many-body approximation schemes~\cite{Hyvarinen:2015bda, Menendez:2017fdf, Deppisch:2020ztt, Ding:2024obt}. Here, we present the relevant set of NMEs for the ground-state-to-ground-state $0^+\rightarrow0^+$ transitions in the $0\nu\beta^+\beta^+, 0\nu\mathrm{EC}\beta^+$ and $0\nu\mathrm{ECEC}$ modes.
\begin{table}[t]
\centering
\renewcommand{\arraystretch}{1.2}
\setlength{\tabcolsep}{5pt}
\begin{tabular}{lrrrrrr}
\hline\hline
NME
&
$^{78}\mathrm{Kr}$
&
$^{96}\mathrm{Ru}$
&
$^{106}\mathrm{Cd}$
&
$^{124}\mathrm{Xe}$
&
$^{130}\mathrm{Ba}$
&
$^{136}\mathrm{Ce}$
\\
\hline
$M_F$
& $-0.678$ & $-0.391$ & $-0.403$ & $-0.403$ & $-0.798$ & $-0.754$ \\
$M_{GT}^{AA}$
& $4.972$ & $4.624$ & $4.159$ & $4.159$ & $4.806$ & $4.661$ \\
$M_{GT}^{AP}$
& $-0.937$ & $-0.872$ & $-0.912$ & $0.982$ & $-0.915$ & $-0.863$ \\
$M_{GT}^{PP}$
& $0.249$ & $0.235$ & $0.259$ & $-0.264$ & $0.246$ & $0.229$ \\
$M_{GT}^{MM}$
& $0.282$ & $0.256$ & $0.278$ & $-0.298$ & $0.278$ & $0.258$ \\
$M_T^{AA}$
& --- & --- & --- & --- & --- & --- \\
$M_T^{AP}$
& $-0.287$ & $0.289$ & $0.293$ & $0.247$ & $-0.219$ & $-0.200$ \\
$M_T^{PP}$
& $0.091$ & $-0.087$ & $-0.095$ & $-0.078$ & $0.070$ & $0.064$ \\
$M_T^{MM}$
& $-0.061$ & $0.058$ & $0.064$ & $0.052$ & $-0.047$ & $-0.043$ \\
$M_{F,\mathrm{sd}}$
& $-1.084$ & $-0.949$ & $-1.042$ & $1.255$ & $-1.175$ & $-1.087$ \\
$M_{GT,\mathrm{sd}}^{AA}$
& $3.749$ & $3.382$ & $3.677$ & $-4.021$ & $3.770$ & $3.502$ \\
$M_{GT,\mathrm{sd}}^{AP}$
& $-1.482$ & $-1.301$ & $-1.458$ & $1.595$ & $-1.494$ & $-1.374$ \\
$M_{GT,\mathrm{sd}}^{PP}$
& $0.478$ & $0.401$ & $0.453$ & $-0.514$ & $-0.482$ & $0.440$ \\
$M_{T,\mathrm{sd}}^{AP}$
& $-0.925$ & $0.868$ & $0.997$ & $0.807$ & $-0.739$ & $-0.678$ \\
$M_{T,\mathrm{sd}}^{PP}$
& $0.367$ & $-0.342$ & $-0.398$ & $-0.321$ & $-0.482$ & $0.271$ \\
\hline\hline
\end{tabular}
\caption{
Nuclear matrix elements for the ground-state-to-ground-state $0^+\rightarrow0^+$ neutrinoless $0\nu\beta^+\beta^+,~0\nu\mathrm{EC}\beta^+$ and $0\nu\mathrm{ECEC}$ modes, calculated in the IBM-2 model.
}
\label{tab:NMEs}
\end{table}

In position-space the relevant long-range NMEs of the light-neutrino-exchange mechanism are defined by~\cite{Cirigliano:2018yza}
\begin{align}
    M_F &= \bra{0^+}\sum_{m,n} h_F(r) \tau^+_{(m)}\tau^+_{(n)}\ket{0^+}, \nonumber
    \\
    M_{GT}^{ij} &= \bra{0^+}\sum_{m,n} h_{GT}^{ij}(r) \boldsymbol{\sigma}_{(m)}\cdot\boldsymbol{\sigma}_{(n)}\tau^+_{(m)}\tau^+_{(n)}\ket{0^+}, \nonumber
    \\
    M_T^{ij} &= \bra{0^+}\sum_{m,n} h_T^{ij}(r) S^{(mn)}(\hat{\mathbf{r}})\tau^+_{(m)}\tau^+_{(n)}\ket{0^+},
\end{align}
with the nuclear form-factors 
\begin{align}
    h_F(\mathbf{q}^2) &= g_V^2(\mathbf{q}^2),\qquad h_{GT}^{AA}(\mathbf{q}^2) = \frac{g_A^2(\mathbf{q}^2)}{g_A^2},\qquad h_{GT}^{AP}(\mathbf{q}^2) = \frac{g_P(\mathbf{q}^2)g_A(\mathbf{q}^2)}{g_A^2}\frac{\mathbf{q}^2}{3m_N},\nonumber
    \\
    h_{GT}^{PP} &= \frac{g_P^2(\mathbf{q}^2)}{g_A^2}\frac{\mathbf{q}^4}{12m_N^2},\qquad h_{GT}^{MM}(\mathbf{q}^2) = \frac{g_M^2(\mathbf{q}^2)}{g_A^2}\frac{\mathbf{q}^2}{6m_N^2},\qquad h_{T}^{AA}(\mathbf{q}^2) = h_{GT}^{AA}(\mathbf{q}^2),\nonumber
    \\
    h_{T}^{AP}(\mathbf{q}^2) &= -h_{GT}^{AP}(\mathbf{q}^2),\qquad h_{T}^{PP}(\mathbf{q}^2) = -h_{GT}^{PP}(\mathbf{q}^2),\qquad h_{T}^{MM}(\mathbf{q}^2) = \frac{1}{2 }h_{GT}^{MM}(\mathbf{q}^2),
\end{align}
entering as~\cite{Hyvarinen:2015bda, Cirigliano:2017djv, Cirigliano:2018yza}
\begin{align}
    h_K^{ij}(r) = \frac{2}{\pi} R_A \int_0^\infty \mathrm{d}|\boldsymbol{q}|\, h_K^{ij}(\mathbf{q}^2) j_\lambda(|\mathbf{q}| r).
\end{align}
Here, $R_A = 1.2\,\mathrm{fm}\times A^{1/3}$ denotes the nuclear radius of the respective isotope while $j_\lambda$ represent the spherical Bessel functions.  For the Fermi and Gamow-Teller transitions $\lambda$ acquires the value $\lambda=0$, while for Tensor transitions we take $\lambda=2$.

The relevant short-distance NMEs of the heavy-neutrino-exchange mechanism can be defined equivalently via~\cite{Cirigliano:2018yza}
\begin{align}
    h_{K,sd}^{ij}(r) &= \frac{2}{\pi} \frac{R_A}{m_\pi^2}\int_0^\infty \mathrm{d}|\boldsymbol{q}|\, \boldsymbol{q}^2 h_K^{ij}(\mathbf{q}^2) j_\lambda(|\mathbf{q}| r),\nonumber
    \\
    M_{F,sd} &= \bra{0^+}\sum_{m,n} h_{F,sd}(r) \tau^+_{(m)}\tau^+_{(n)}\ket{0^+}, \nonumber
    \\
    M_{GT,sd}^{ij} &= \bra{0^+}\sum_{m,n} h_{GT,sd}^{ij}(r) \boldsymbol{\sigma}_{(m)}\cdot\boldsymbol{\sigma}_{(n)}\tau^+_{(m)}\tau^+_{(n)}\ket{0^+}, \nonumber
    \\
    M_{T,sd}^{ij} &= \bra{0^+}\sum_{m,n} h_{T,sd}^{ij}(r) S^{(mn)}(\hat{\mathbf{r}})\tau^+_{(m)}\tau^+_{(n)}\ket{0^+}.
\end{align}
In Table~\ref{tab:NMEs} we present the relevant NMEs for the naturally occurring $0\nu\beta^+\beta^+$ candidate isotopes calculated using the IBM-2 model. 

\subsection{Phase Space Factors}\label{sec:PSFs}
The relevant PSFs for the neutrinoless decay modes depend on the same positron wavefunctions as the neutrinoful scenario described in Sec.~\ref{sec:2nuPSFs}.

\subsubsection{$0\nu\beta^+\beta^+$}
\begin{table}[t]
\centering
\renewcommand{\arraystretch}{1.2}
\setlength{\tabcolsep}{6pt}
\begin{tabular}{lrrrrrr}
\hline\hline
PSF
&
$^{78}\mathrm{Kr}$
&
$^{96}\mathrm{Ru}$
&
$^{106}\mathrm{Cd}$
&
$^{124}\mathrm{Xe}$
&
$^{130}\mathrm{Ba}$
&
$^{136}\mathrm{Ce}$
\\
\hline
$G_{01}$ & $279.53$ & $94.82$ & $107.88$ & $127.08$ & $28.97$ & $2.78$ \\
$G_{02}$ & $41.91$  & $8.78$   & $12.21$  & $18.67$  & $1.74$  & $0.04$ \\
$G_{03}$ & $39.95$  & $9.77$  & $12.06$  & $15.66$  & $2.10$  & $0.08$ \\
$G_{04}$ & $195.79$ & $63.46$ & $75.46$ & $94.03$ & $19.13$ & $1.49$ \\
$G_{06}$ & $627.43$ & $223.01$ & $243.02$ & $268.72$ & $69.14$ & $7.64$ \\
$G_{09}$ & $728.87$ & $253.48$ & $282.23$ & $322.93$ & $78.17$ & $8.16$ \\
\hline
$Q~[\mathrm{MeV}]$ & $0.802$ & $0.670$ & $0.731$ & $0.820$ & $0.575$ & $0.335$ \\
\hline\hline
\end{tabular}
\caption{
Phase-space factors for the $0\nu\beta^+\beta^+$ modes in units of
$10^{-20}\,\mathrm{yr}^{-1}$, calculated using the \textit{``exact''} numerical solution to the electron/positron wavefunctions in a screened Coulomb potential with a uniformly charged nucleus.
The last row gives the corresponding $Q$ values in MeV.
}
\label{tab:PSFs}
\end{table}
The PSFs of the $0\nu\beta^+\beta^+$ decay can be written in the general form~\cite{Stefanik:2015twa} as
\begin{align}
    G_{0k}^{\beta^+\beta^+} =& C_{0k}\frac{G_F^4m_e^2}{64\pi^5\ln{2}R^2}\int \delta\bigg(\epsilon_1 + \epsilon_2 +E_f-E_i\bigg) \nonumber
    \\
    &\times \bigg(h_{0k}(\epsilon_1, \epsilon_2, R)\cos{\theta} + g_{0k}(\epsilon_1, \epsilon_2, R)\bigg) k_1 k_2 \epsilon_1 \epsilon_2 \,\text{d}\epsilon_1\,\text{d}\epsilon_2\,\text{d}(\cos{\theta}),\label{eq:PSFs}
\end{align}
where the functions $h_{0k},g_{0k}$ parameterize the angular and radial part of the leptonic current calculated from the corresponding electron/positron wave functions, $\epsilon_i, k_i$ represent the energy and momentum of the $i-$th electron/positron, $\theta$ is the opening angle between the two emitted leptons, and $E_i, E_f$ are the energy of the initial and final state nucleus, respectively. 
The prefactors $C_{0k}$ take care of a convention-dependent rescaling between Refs.~\cite{Doi:1992dm, Stefanik:2015twa} and the EFT approach of Refs.~\cite{Cirigliano:2017djv, Cirigliano:2018yza}, which we follow. They are given as
\begin{align}
    C_{04} = \frac{9}{2},\qquad C_{06} = \frac{m_e R}{2},\qquad C_{09} = \left(\frac{m_e R}{2}\right)^2~~\text{and}~~ C_{0k}=1~\text{otherwise}.\label{eq:prefactors}
\end{align}
The relevant set of PSFs for $0\nu\beta^+\beta^+$ is given in Table~\ref{tab:PSFs}. We refer the reader to Ref.~\cite{Stefanik:2015twa} for a more detailed discussion of the PSF calculation process and the precise wavefunction-dependent definitions of $g_{0k}$ and $h_{0k}$.

\subsubsection{$0\nu\mathrm{EC}\beta^+$}
In the neutrinoless version of the mixed capture and emission mode, $0\nu\mathrm{EC}\beta^+$, the energies of the bound-state electron $\epsilon_{e}$ and the emitted positron $\epsilon_{p}$ are fixed as
\begin{align}
    \epsilon_{e,n} = m_e - \epsilon_{b,n},\qquad \epsilon_{p,n} = m_e + Q - \epsilon_{b,n} , \qquad Q_{\mathrm{EC\beta^+}} = \Delta M - 2m_e.
\end{align}
Consequently, the radial wavefunctions only need to be evaluated at their respective energies and no integral needs to be solved. 

The overall PSFs are then given by~\cite{Kotila:2013gea, Doi:1992dm}
\begin{align}
    G_{0k}^{\mathrm{EC}\beta^+} = C_k \frac{G_F^4 m_e^5}{8\log(2)\pi^3} \times\sum_n\mathcal{B}_{n,-1}^2 g_{0k}^{\mathrm{EC}\beta^+}k_{p}\epsilon_{p},
\end{align}
with~\cite{Doi:1992dm}
\begin{align}
    g_{01}^{\mathrm{EC}\beta^+}
        &= g_{-1}^2(\epsilon_{p},R) + f_{+1}^2(\epsilon_{p},R),
    &
    g_{02}^{\mathrm{EC}\beta^+}
        &= \left(\frac{\epsilon_{12}}{m_e}\right)^2
           g_{-1}^2(\epsilon_{p},R),\nonumber
    \\
    g_{03}^{\mathrm{EC}\beta^+}
        &= 2\frac{\epsilon_{12}}{m_e}
           g_{-1}^2(\epsilon_{p},R),
    &
    g_{04}^{\mathrm{EC}\beta^+}
        &= \frac{4}{9} f_e\, g_{-1}^2(\epsilon_{p},R),\nonumber
    \\
    g_{06}^{\mathrm{EC}\beta^+}
        &= \frac{8}{m_e R} f_{+1}^2(\epsilon_{p},R),
    &
    g_{09}^{\mathrm{EC}\beta^+}
        &= \left(\frac{4}{m_e R}\right)^2
           f_{+1}^2(\epsilon_{p},R),
\end{align}
and $f_e = 1-3\alpha/(2m_e R)$. After applying the convention-dependent prefactors $C_{0k}$ defined in eq.~\eqref{eq:prefactors}, the two PSFs $G_{06}$ and $G_{09}$ turn out to be equal in the leading order approximation of Ref.~\cite{Doi:1992dm}. The resulting PSFs for the $0\nu\mathrm{EC}\beta^+$ process are displayed in Table~\ref{tab:PSFsECb+}. Compared to the PSFs of the double positron emitting $0\nu\beta^+\beta^+$ decay, the $0\nu\mathrm{EC}\beta^+$ PSFs are much less sensitive to the chosen isotope as the relative difference in $Q-$values is much smaller.

\begin{table}[t]
\centering
\renewcommand{\arraystretch}{1.2}
\setlength{\tabcolsep}{6pt}
\begin{tabular}{lrrrrrr}
\hline\hline
PSF
&
$^{78}\mathrm{Kr}$
&
$^{96}\mathrm{Ru}$
&
$^{106}\mathrm{Cd}$
&
$^{124}\mathrm{Xe}$
&
$^{130}\mathrm{Ba}$
&
$^{136}\mathrm{Ce}$
\\
\hline
$G_{01}$ & $6.81$ & $10.36$ & $14.02$ & $21.37$ & $19.08$ & $16.57$ \\
$G_{02}$ & $125.97$ & $173.71$ & $243.10$ & $387.69$ & $292.33$ & $212.21$ \\
$G_{03}$ & $45.64$ & $66.37$ & $91.09$ & $141.38$ & $117.07$ & $94.03$ \\
$G_{04}$ & $1.45$ & $2.93$ & $4.37$ & $7.57$ & $7.14$ & $6.57$ \\
$G_{06}$ & $10.70$ & $16.06$ & $21.97$ & $33.93$ & $29.42$ & $24.62$ \\
$G_{09}$ & $10.70$ & $16.06$ & $21.97$ & $33.93$ & $29.42$ & $24.62$ \\
\hline
$Q~[\mathrm{MeV}]$ & $1.824$ & $1.692$ & $1.753$ & $1.842$ & $1.596$ & $1.356$ \\
\hline\hline
\end{tabular}
\caption{
Phase-space factors for the $0\nu\mathrm{EC}\beta^+$ modes in units of
$10^{-18}\,\mathrm{yr}^{-1}$.
}
\label{tab:PSFsECb+}
\end{table}

\subsubsection{$0\nu\mathrm{ECEC}$}
While the double electron capture mode is the most sensitive decay channel in the neutrino-emitting mode due to its larger phase-space, the situation is much more convoluted in the neutrinoless case. This is simply because in the neutrinoless double electron capture, no particles are emitted by the nucleus. Therefore, energy conservation requires the initial and final state to be degenerate, thus resulting in a resonant mechanism for the $0\nu\mathrm{ECEC}$ half-life~\cite{Bernabeu:1983yb, Krivoruchenko:2010ng, Kotila:2014zya, Blaum:2020ogl} (although it should be noted that alternative non-resonant mechanisms have been proposed as well~\cite{Karpeshin:2020fad}). Within the standard picture, treating $0\nu\mathrm{ECEC}$ as a resonant process, the half-life can be expressed in analogy to eq.~\eqref{eq:0nuhalflife} as
\begin{align}
    \left(T_{1/2}^{0\nu\mathrm{ECEC}}\right)^{-1} = g_A^4 \sum_k G_{0k}^\mathrm{ECEC}|\mathcal{A}_k|^2\times\frac{m_e\Gamma}{\Delta^2 + \Gamma^2/4},
\end{align}
where the last term encodes the resonant behaviour in terms of the degeneracy parameter $\Delta$ describing the energy difference between the initial and the final states and the decay width of the final state $\Gamma=\Gamma_{h_1} + \Gamma_{h_2} + \Gamma^*$ in terms of the individual electron hole widths $\Gamma_{h_i}$ as well as the de-excitation width of the final state nucleus $\Gamma^*$ in a ground-state-to-excited-state transition, the latter of which can usually be neglected~\cite{Kotila:2014zya}. With typical decay widths of the order of a few eV~\cite{Blaum:2020ogl} the identification of potential resonant decay candidates requires extremely precise determination of the initial and the final state energies. Reference~\cite{Blaum:2020ogl} provides a comprehensive overview of possible $0\nu\mathrm{ECEC}$ candidate isotopes. Due to the associated large uncertainty involved in the half-life calculation, with half-life estimates often spanning over 6-7 orders of magnitude in the most promising candidate isotopes, we refrain from a more detailed analysis of this decay mode for now.

\subsection{Spectra and Angular Correlation}\label{sec:phase_space_observables}
If individual electron/positron kinematics can be experimentally resolved, both the energy spectra of the individually emitted leptons as well as their angular correlation in the $0\nu\beta\beta$ decay may provide further insight into the underlying LNV physics~\cite{Graf:2022lhj}.
With the PSFs defined in eq.~\eqref{eq:PSFs}, the single electron/positron spectra in the $0\nu\beta^\pm\beta^\pm$ decays take the form~\cite{Stefanik:2015twa, Graf:2022lhj},
\begin{align}
    \frac{\mathrm{d}\Gamma}{\mathrm{d}\epsilon_1} = \frac{G_F^4 m_e^2}{32\pi^5 R_A^2} \sum_k C_kg_{0k}|\mathcal{A}_k|^2 k_1k_2\epsilon_1\epsilon_2,
\end{align}
which is, again, conveniently normalized as
\begin{align}
    \frac{1}{\Gamma}\frac{\mathrm{d}\Gamma}{\mathrm{d}\overline{\epsilon}_1} = \frac{Q}{\Gamma}\frac{\mathrm{d}\Gamma}{\mathrm{d}{\epsilon}_1},\qquad \overline{\epsilon}_1 = \frac{\epsilon_1-m_e}{Q}.
\end{align}

Similarly, we can describe the angular correlation of the two emitted electrons/positrons with the angular correlation coefficient given by
\begin{align}
    \frac{a_1}{a_0} = \frac{\sum_i|\mathcal{A}_i|^2h_{0i}}{\sum_j|\mathcal{A}_j|^2g_{0j}}.
\end{align}
Generally, in single-operator dominated scenarios, there are three classes of qualitatively different signatures to be expected.
First, the \textit{standard} \Lnuem~dominated by the Majorana neutrino mass shows a negative angular correlation with a single electron/positron spectrum peaked at $\overline{\epsilon}=0.5$, i.e.,~the two electrons/positrons are preferably emitted back-to-back with similar energies. When considering LNV SMEFT dimension-7 operators, most of them share the same qualitative phase-space signatures as \Lnuem (either at tree level or by mixing with the Weinberg operator at the 1-loop level~\cite{Zhang:2023kvw, Graf:2025cfk}). The exceptions are the vector operator $\mathcal{O}_{LHDe}^{(7)}$, which features the same spectrum as \Lnuem, but gives a positive angular correlation, and $\mathcal{O}_{LeudH}^{(7)}$, which results in a spectrum that favours an uneven energy distribution between electrons and an angular correlation that is positive for uneven electron/positron energies and negative around $\overline{\epsilon}=0.5$.
\begin{figure}[t!]
    \centering
    \includegraphics[width=\linewidth]{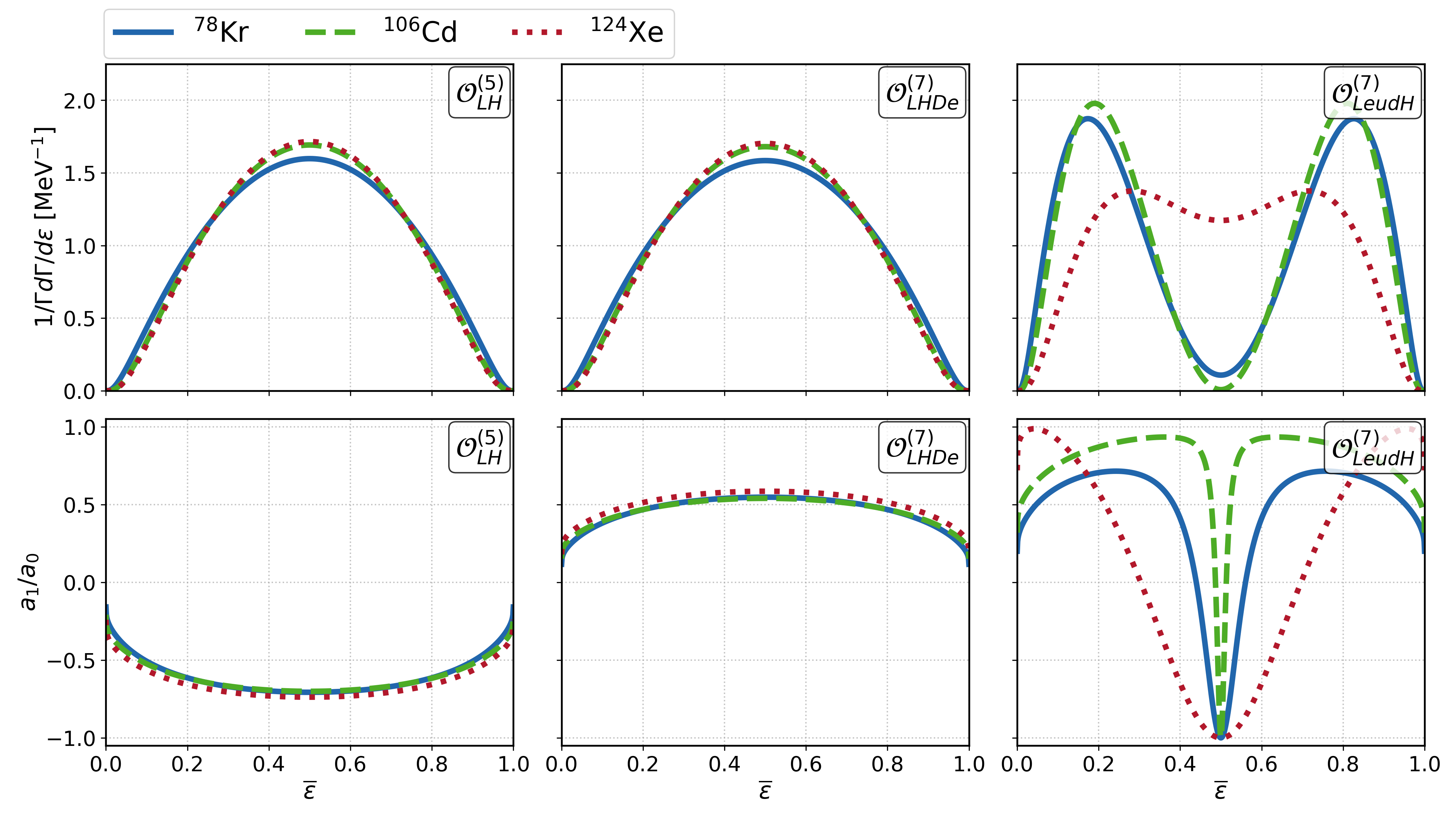}\vspace{-1cm}
    \caption{Normalized single positron spectra (upper) and angular correlation coefficients (lower). The phase-space observables are shown for the three $0\nu\beta^+\beta^+$ candidate isotopes of NuDoubt++ (\textsuperscript{78}Kr, \textsuperscript{106}Cd, \textsuperscript{124}Xe). Only the signatures of operators $\mathcal{O}_{LHDe}^{(7)}$ and $\mathcal{O}_{LeudH}^{(7)}$ are distinct from the light-neutrino-exchange mechanism corresponding to the Weinberg operator $\mathcal{O}_{LH}^{(5)}$.}
    \label{fig:spectra}
\end{figure}

In Figure~\ref{fig:spectra} we show normalized single-positron spectra and angular correlations for $0\nu\beta^+\beta^+$ decay triggered by the different dimension-5 and dimension-7 LNV SMEFT operators in candidate isotopes \textsuperscript{78}Kr, \textsuperscript{106}Cd, and \textsuperscript{124}Xe.

\subsection{Projected Sensitivity of NuDoubt++}\label{sec:BSM_sensitivity}
\subsubsection{Single Operator Dominance}
\begin{figure}[t!]
    \centering
    \includegraphics[width=\linewidth]{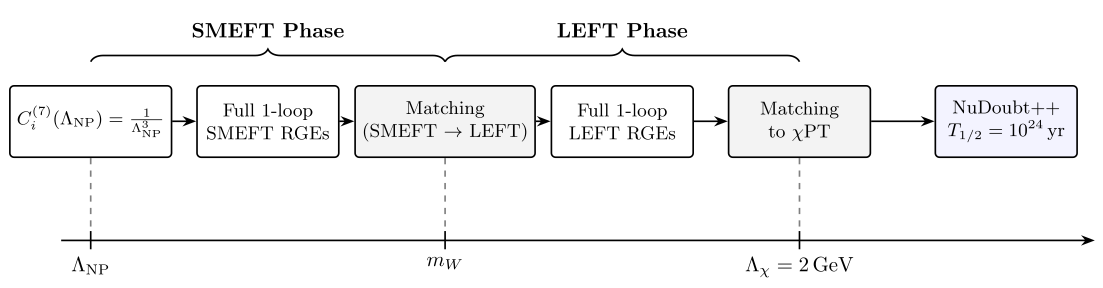}
    \vspace{-1cm}
    \caption{Schematic visualization of the running and matching procedure applied to derive limits on the new-physics scale $\Lambda_\mathrm{NP}$. The procedure is solved iteratively until the new physic scale $\Lambda_\mathrm{NP}$ matches the projected experimental sensitivity of $T_{1/2}=10^{24}\,\mathrm{yr}$.}
    \label{fig:limits_extraction}
\end{figure}

By applying a novel hybrid opaque loaded scintillator concept~\cite{LiquidO:2019mxd, Buck:2019tsa, Steiger:2024nes}, the NuDoubt++ collaboration~\cite{NuDoubt:2024jax} aims to achieve the first experimental detection of positron emitting double beta decay. The primary candidate isotope of the experiment is \textsuperscript{78}Kr, with the potential to also study \textsuperscript{106}Cd and \textsuperscript{124}Xe in later phases. With an initial targeted exposure of $1\,\mathrm{tonne\cdot week}\sim20\,\mathrm{kg\cdot yr}$ with respect to the scintillator mass, NuDoubt++ is predicted to observe the $2\nu\mathrm{EC}\beta^+$ process, while simultaneously exploring the neutrinoless $0\nu\beta^+\beta^+$ decay mode up to a half-life limit of $10^{24}\,\mathrm{yr}$~\cite{NuDoubt:2024jax}. Here, we study the projected exclusion sensitivity of the NuDoubt++ experiment towards LNV dimension-7 SMEFT operators. By taking
\begin{align}
    C_{LH}^{(5)} = \frac{1}{\Lambda_\mathrm{NP}},\qquad C_i^{(7)} = \frac{1}{\Lambda_\mathrm{NP}^3},\label{eq:WCscale}
\end{align}
we can interpret the corresponding limits on the dimension-5 Weinberg operator as well as the dimension-7 Wilson coefficients in terms of a new-physics scale $\Lambda_\mathrm{NP}$. We calculate the projected limits on the corresponding new-physics scale $\Lambda_\mathrm{NP}$ by iteratively solving the full 1-loop RGE running and matching process down to the SMEFT-to-LEFT matching scale $m_W$ and the subsequent RGE evolution and matching down to the chiral matching scale $\Lambda_\chi=2\,\mathrm{GeV}$ until the resulting half-life matches the NuDoubt++ projected half-life sensitivity of $T_{1/2}=10^{24}\,\mathrm{yr}$. See Figure~\ref{fig:limits_extraction} for a visualization of the iterative running and matching procedure. The significant importance of including loop effects in \0 calculations was recently shown in Ref.~\cite{Graf:2025cfk}.

\begin{figure}[t!]
    \centering
    \includegraphics[width=1\linewidth]{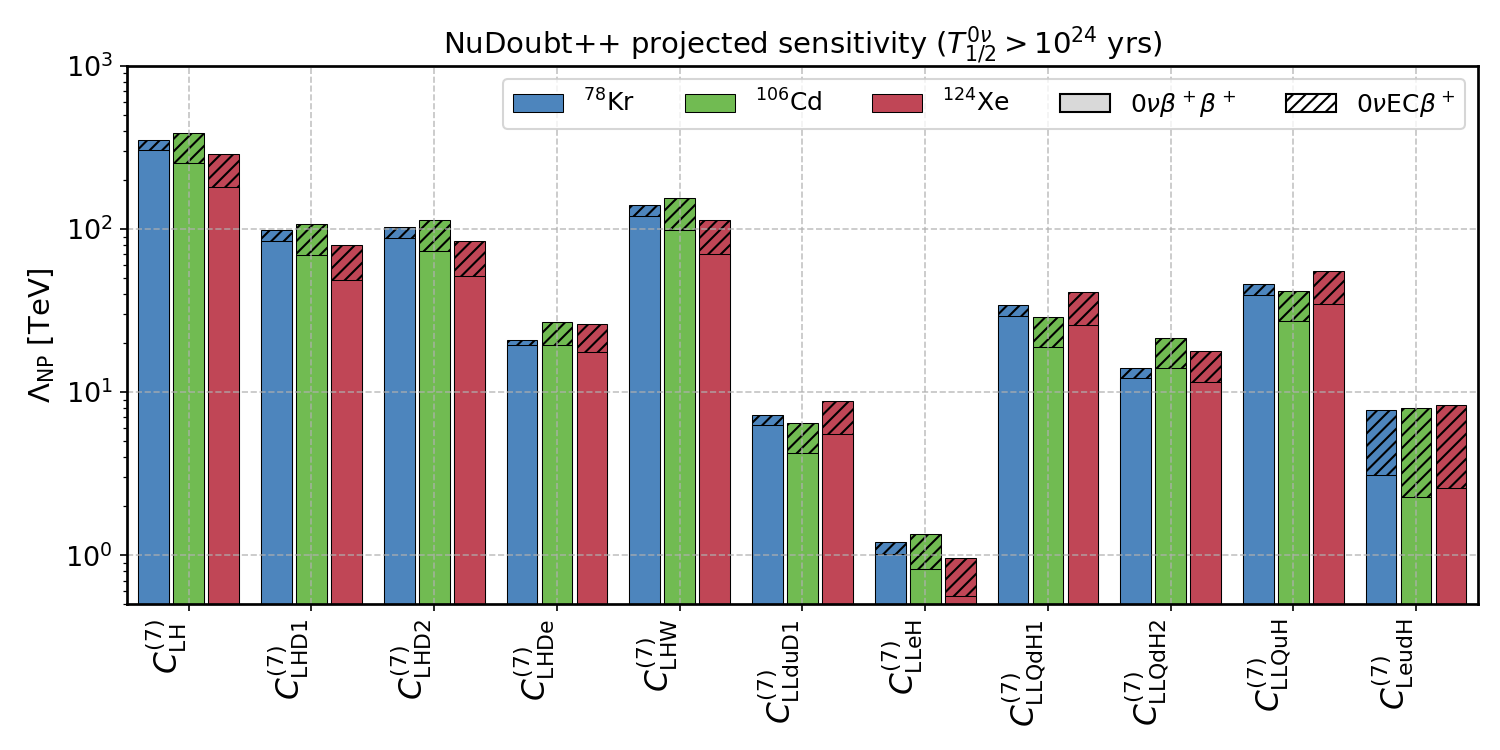}\vspace{-1cm}
    \caption{Projected exclusion reach of the NuDoubt++ experiment for the new-physics scale $\Lambda_\mathrm{NP}$ associated with the lepton-number-violating dimension-7 SMEFT operators inducing $0\nu\beta^+\beta^+$ decay.}
    \label{fig:smeft_limits}
\end{figure}

In Figure~\ref{fig:smeft_limits} we display the projected exclusion sensitivity of the NuDoubt++ experiment in terms of the corresponding new physics scale $\Lambda_\mathrm{NP}$ for the relevant LNV dim-7 SMEFT operators. Despite the unfavorable phase-space suppression, the projected new physics scale sensitivity of the NuDoubt++ experiment is in the range of $\mathcal{O}(1)\,\mathrm{TeV} - \mathcal{O}(10^2)\,\mathrm{TeV}$ and thereby within one order of magnitude to the most competitive $0\nu\beta^-\beta^-$ limits provided by the KamLAND-Zen collaboration~\cite{KamLAND-Zen:2024eml, Graf:2025cfk}. The comparable sensitivity to the supposed new physics scale $\Lambda_\mathrm{NP}$ despite the significant difference in half-life sensitivities between the NuDoubt++ projection and current state-of-the-art $0\nu\beta^-\beta^-$ experiments can be attributed to the strong scale dependence of the \0 half-life at SMEFT dimension 7 of $T^{0\nu}_{1/2}\propto \Lambda_\mathrm{NP}^6$ or conversely $\Lambda_\mathrm{NP}\propto T_{1/2}^{0\nu \ 1/6}$.

For the dimension-5 Weinberg operator, the projected exclusion sensitivity is in the range of $\Lambda_\mathrm{NP}\sim\mathcal{O}(10^{8-9})\,\mathrm{TeV}$ corresponding to an effective Majorana neutrino mass of $m_{\beta\beta}\sim\mathcal{O}(10-100)\,\mathrm{eV}$, scaling as $\Lambda_\mathrm{NP}\propto T_{1/2}^{0\nu \ 1/2}$

Beyond the simple sensitivity projection, the mechanism-dependent ratio of the $0\nu\beta^+\beta^+$ and $0\nu\mathrm{EC}\beta^+$ half-lives provides a potential discriminator: observing both modes could help confirm or exclude a right-handed-current scenario induced by the $\mathcal{O}^{(7)}_{LeudH}$ operator.
\begin{figure}[t!]
    \centering
    \includegraphics[width=\linewidth]{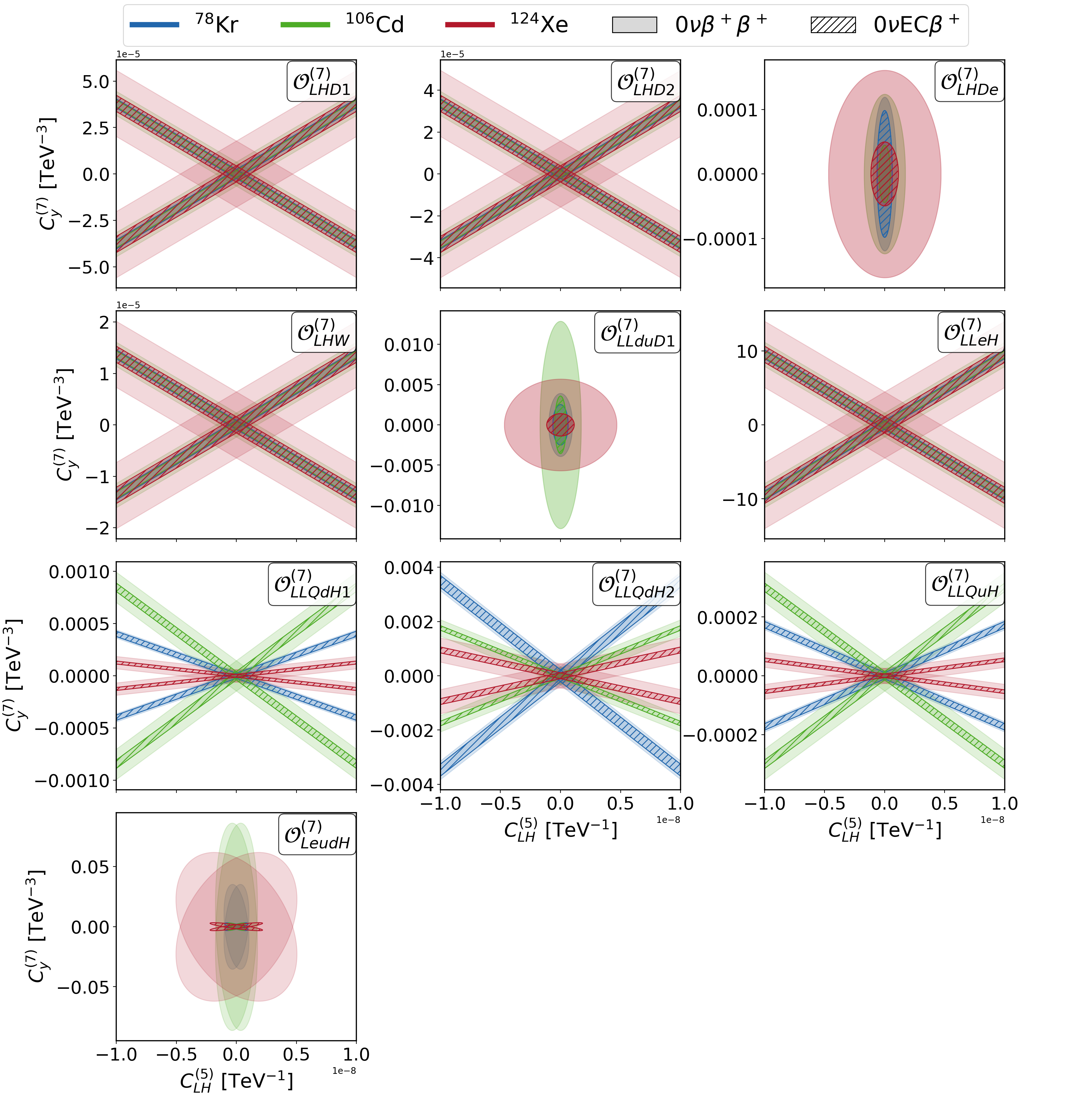}\vspace{-0.5cm}
    \caption{Projected $C_i^{(7)}-C_{LH}^{(5)}$ parameter space that can be constrained by the future NuDoubt++ experiment using three different $0\nu\beta^+\beta^+$ candidate isotopes \textsuperscript{78}Kr, \textsuperscript{106}Cd, and \textsuperscript{124}Xe}
    \label{fig:contours}
\end{figure}

\subsubsection{Two-Operator Scenarios}
Assuming that the $0\nu\beta^+\beta^+$ decay rate receives dominant contributions from two EFT operators, $\mathcal{O}_x$ and $\mathcal{O}_y$, the corresponding inverse half-life takes the general quadratic form that defines an ellipse in the associated parameter space~\cite{Scholer:2023bnn},
\begin{align}
    T_{1/2}^{0\nu} = M_{xx}|C_{x}|^2 + M_{yy}|C_{y}|^2 + 2M_{xy}\mathrm{Re}\Big[ C_x C_y \exp{(i\theta)}\Big]\,,
\end{align}
where $|C_{x/y}|$ denotes the magnitude of the corresponding Wilson coefficient, while $\theta$ is the relative complex phase between the two coefficients. In such a scenario, strong destructive interference between the two operator contributions can generate nearly unconstrained directions in the $C_x$--$C_y$ parameter space. Since the allowed regions inferred from $0\nu\beta^+\beta^+$ measurements in different isotopes can be tilted relative to one another, combining measurements from multiple isotopes can lift these degeneracies and exclude otherwise weakly constrained directions.

In Figure~\ref{fig:contours}, we show the projected constraints in the $C_{LH}^{(5)}$--$C_y^{(7)}$ parameter space arising from the interference between the Weinberg operator and the dimension-7 LNV SMEFT operator $\mathcal{O}_y^{(7)}$. Results are presented for the three isotopes that could be deployed in the NuDoubt++ experimental setup. To account for renormalization-group evolution, we assume that the operators are generated at a scale of $1,\mathrm{TeV}$.

For the operators $\mathcal{O}_{LLQdH1}$, $\mathcal{O}_{LLQdH2}$, and $\mathcal{O}_{LLQuH}$, strong destructive interference gives rise to nearly unconstrained directions in the $C_{LH}^{(5)}$--$C_y^{(7)}$ parameter space. These degeneracies can be lifted by combining measurements of the $0\nu\beta^+\beta^+$ half-life in two or more of the candidate NuDoubt++ isotopes. By contrast, the operators $\mathcal{O}_{LHDe}$, $\mathcal{O}_{LLduD1}$, and $\mathcal{O}_{LeudH}$ lead to simple elliptical allowed regions, without pronounced destructive interference with the contribution induced by the Weinberg operator. Nevertheless, combining half-life measurements across multiple isotopes and decay modes can further tighten the allowed regions in the $C_{LH}^{(5)}$--$C_y^{(7)}$ parameter space.

As expected, the operators $\mathcal{O}_{LHD1}$, $\mathcal{O}_{LHD2}$, $\mathcal{O}_{LHDe}$, $\mathcal{O}_{LHW}$, and $\mathcal{O}_{LLeH}$, which mix strongly with the dimension-5 Weinberg operator and its dimension-7 analogue under renormalization-group evolution~\cite{Zhang:2023kvw,Zhang:2023ndw,Graf:2025cfk}, exhibit nearly unconstrained directions with little relative tilt among the contours obtained for different isotopes. Owing to this strong mixing with $\mathcal{O}_{LH}^{(5,7)}$, the corresponding contour plots effectively probe the $C_{LH}^{(5,7)}$--$C_{LH}^{(5)}$ parameter space. By construction, this plane contains weakly constrained directions associated with destructive interference between the two contributions.

\section{Conclusion}

In this work we have studied positron-emitting and electron-capturing double-beta-decay modes in the context of the proposed NuDoubt++ experimental program. While conventional double-beta-decay searches focus primarily on the $\beta^-\beta^-$ channel, the complementary $\beta^+\beta^+$, $\beta^+\mathrm{EC}$, and ECEC modes provide an independent handle on both the nuclear-structure calculations involved and the possible lepton-number-violating physics. We have presented a dedicated analysis of the NuDoubt++ candidate isotopes $^{78}\mathrm{Kr}$, $^{106}\mathrm{Cd}$, and $^{124}\mathrm{Xe}$, combining Standard-Model two-neutrino estimates with an effective-field-theory treatment of neutrinoless decay induced by dimension-seven SMEFT operators.

For the Standard-Model modes, we have collected the relevant IBM-2 nuclear matrix elements, updated the phase-space factors and derived the corresponding half-life estimates. The electron-capture channels are the most promising near-term targets, with $2\nu\mathrm{ECEC}$ and $2\nu\beta^+\mathrm{EC}$ potentially accessible in NuDoubt++-like exposures\footnote{Though, a $2\nu\mathrm{ECEC}$ discovery would require an updated setup with improved sensitivity in the relevant energy regime compared to the initial NuDoubt++ proposal~\cite{stefan}.}. By contrast, the fully positron-emitting $2\nu\beta^+\beta^+$ mode is strongly phase-space suppressed. In a simplified single-bin low-background analysis we find that its observation would require approximately $10^2$--$10^4$ tonne $\times$ days of $^{78}\mathrm{Kr}$ exposure, depending on the true half-life and background assumptions. This estimate should be interpreted as indicative, since a realistic detector-level treatment of the full $2\nu\beta^+\beta^+$ spectrum and backgrounds may shift the required exposure.

For the neutrinoless modes, we have computed the relevant $0\nu\beta^+\beta^+$ and $0\nu\mathrm{EC}\beta^+$ phase-space factors and nuclear matrix elements and embedded them in the EFT master-formula approach. Interpreting the projected NuDoubt++ sensitivity $T_{1/2}^{0\nu}=10^{24}\,\mathrm{yr}$ in terms of dimension-seven SMEFT operators, and assuming $C_i^{(7)}=1/\Lambda_\mathrm{NP}^3$, we find that NuDoubt++ can probe new-physics scales in the range $\mathcal{O}(1)$--$\mathcal{O}(10^2)\,\mathrm{TeV}$. Although the positron-emitting channel suffers from reduced phase space compared to standard $0\nu\beta^-\beta^-$ searches, the resulting reach is only moderately weaker at the level of the inferred SMEFT scale because of the mild scaling $\Lambda_\mathrm{NP}\propto T_{1/2}^{0\nu \ 1/6}$.

We have also shown that the ability to study several candidate isotopes within the same experimental concept is particularly useful in scenarios with more than one active lepton-number-violating operator. In such cases, destructive interference can create approximately unconstrained directions in Wilson-coefficient parameter space. Since the corresponding allowed regions are isotope dependent, combined information from $^{78}\mathrm{Kr}$, $^{106}\mathrm{Cd}$, and $^{124}\mathrm{Xe}$ can help break these degeneracies and improve the interpretation of a future signal or limit. This makes NuDoubt++ not only a search experiment for rare positron-emitting double-beta modes, but also a potentially valuable diagnostic tool for the underlying mechanism of lepton number violation. 

In principle, the neutrinoless double electron capture mode ($0\nu\mathrm{ECEC}$) may offer similar benefits as the study of positron-emitting decay modes. However, half-life estimates and corresponding projected limits on new-physics parameters suffer from significant numerical uncertainties related to the nature of the resonance mechanism. Hence, a robust study of this mode is beyond the current feasibility.

The main results of this work can be summarized as follows:
\begin{itemize}
    \item We provide IBM-2 nuclear matrix elements and accurate phase-space factors for the relevant $2\nu$ and $0\nu$ positron-emitting and electron-capturing modes in all naturally occuring $0\nu\beta^+\beta^+$ candidate isotopes; \textsuperscript{78}Kr, \textsuperscript{96}Ru, \textsuperscript{106}Cd, \textsuperscript{124}Xe, \textsuperscript{130}Ba, and \textsuperscript{136}Ce.

    \item The $2\nu\mathrm{ECEC}$ and $2\nu\beta^+\mathrm{EC}$ modes are the most promising Standard-Model channels when considering comparatively low-exposure kg-scale scenarios like the NuDoubt++ proposal.

    \item Observation of $2\nu\beta^+\beta^+$ is substantially more challenging and may require $10^2$--$10^4$ tonne $\times$ days of $^{78}\mathrm{Kr}$ exposure in a simplified low-background analysis, suggesting the necessity for experiments in the $\mathcal{O}(100\,\mathrm{kg})-\mathcal{O}(1000\,\mathrm{kg})$ regime with respect to the total isotope mass.

    \item For a projected $0\nu\beta^+\beta^+$ half-life sensitivity of $10^{24}\,\mathrm{yr}$, NuDoubt++ can probe dimension-seven LNV SMEFT operators corresponding to new-physics scales approximately of the order of $\mathcal{O}(1)$--$\mathcal{O}(10^2)\,\mathrm{TeV}$.

    \item Phase-space observables distinguish only a small subset of SMEFT operators, most notably $O^{(7)}_{LHDe}$ and $O^{(7)}_{LeudH}$, although the experimental accessibility of these observables in NuDoubt++ should be assessed with a dedicated detector-level study.

    \item Multi-isotope measurements can help lift degeneracies in two-operator scenarios by constraining destructive-interference directions in Wilson-coefficient space. Positron-emitting double-beta searches therefore provide a complementary probe of LNV physics, with interpretation power that is not captured by half-life sensitivity alone. NuDoubt++ offers the unique opportunity to test multiple $0\nu\beta\beta$ candidate isotopes within the same experimental setup. This allows to study unconstrained regions of destructive interference in the parameter space of multi-operator scenarios.
\end{itemize}

\section*{Acknowledgements}
We thank Stefan Schoppmann for valuable comments on the manuscript in the context of its relevance to the proposed NuDoubt++ experiment.
L.~G. acknowledges support from the Dutch Research Council (NWO) under project number VI.Veni.222.318 and from Charles University through the project number PRIMUS/24/SCI/013. O. S. acknowledges support by the Alexander von Humboldt Foundation under the Feodor Lynen Research Fellowship program and by the National Science Foundation under cooperative agreement 2020275. J.K. acknowledges
support from project PNRR-I8/C9-CF264, Contract No. 760100/23.5.2023 of the Romanian
Ministry of Research, Innovation and Digitalization (the NEPTUN project).


\bibliographystyle{JHEP}
\bibliography{references}
\end{document}